\documentclass[iop,apj]{emulateapj}
\usepackage{amssymb}
\usepackage[english]{babel}
\usepackage{hyperref}

\usepackage{amsmath,amssymb,amsxtra,amsfonts}
\usepackage{graphicx}
\newcommand{\bea}{\begin{eqnarray}}
\newcommand{\eea}{\end{eqnarray}}
\newcommand\be{\begin{equation}}
\newcommand\ee{\end{equation}}

\shorttitle{Power of the redshift drift} \shortauthors{Zhang et al.}

\begin{document}

\title{Power of the redshift drift on cosmological models and expansion history}

\author{Ming-Jian Zhang \altaffilmark{1},  and Wen-Biao Liu \altaffilmark{1,*}}
\email{wbliu@bnu.edu.cn}

\altaffiltext{1}{Department of Physics, Institute of Theoretical
Physics, Beijing Normal University, Beijing, 100875, China}

\begin{abstract}
We investigate the power of the velocity drift ($\Delta v$) on
cosmological parameters and expansion history with observational
Hubble data (OHD), type Ia supernova (SNIa). We estimate the
constraints of $\Delta v$  using the Fisher information matrix based
on the model by \citet{pasquini2005codex,whitelock2006scientific}.
We find that $\Delta v$ with 20 years can reduce the uncertainty of
$\Omega_m$ by more than 42\% than available observations. Based on
the statistical figures of merit (FoM), we find that in order to
match the constraint power of OHD and SNIa, we need 21 and 26 future
measurements, respectively. We also quantitatively estimate for the
first time the number of years required for the velocity drift to
become comparable with current observations on the equation of state
$w$. The statistical FoM indicate that we need at least 12 years to
cover current observations. Physically, we could monitor 30 quasars
for 30 years to obtain the same accuracy of $w$. Considering two
parameterized deceleration factor $q(z)$, we find that the available
observations give an estimation on current value $-0.9 \lesssim q_0
\lesssim -0.3$. Difference between the two types of $q(z)$ is the
precise determination of variation rate $dq/dz$. For the first model
with constant $dq/dz$, $\Delta v$ with only 10 years provides a much
better constraint on it, especially when compared with SNIa.
However, we need $\Delta v$ for more years in the variable $dq/dz$
model. We find that $\Delta v$ with 30 years reduces the uncertainty
of transition redshift to approximately three times better than
those of OHD and SNIa.

\end{abstract}

\keywords{cosmology: dark energy --- deceleration factor ---  Fisher
matrix --- velocity drift   }

\section{Introduction}
\label{introduction}

Cosmological observations probe an accelerating expansion of the
universe.  Examples include Type Ia supernova (SNIa) observations
\citep{riess1998supernova}, Large Scale Structure
\citep{tegmark2004cosmological}, and Cosmic Microwave Background
(CMB) anisotropy \citep{spergel2003first}.  As a key factor that
reflects the expansion history of the universe, the Hubble parameter
$H=\dot{a}/a$ relevant to various observations. In practice, we
measure the Hubble parameter as a function of redshift $z$.
Observationally, we can deduce $H(z)$ from differential ages of
galaxies
\citep{jimenez2008constraining,simon2005constraints,stern2010cosmic},
from the baryon acoustic oscillation (BAO) peaks in the galaxy power
spectrum \citep{gaztanaga2009clustering,moresco2012improved} or from
the BAO peak using the Ly$\alpha$ forest of quasars
\citep{2013A&A...552A..96B}. We can theoretically reconstruct $H(z)$
from the luminosity distances of SNIa using their differential
relations
\citep{wang2005uncorrelated,shafieloo2006smoothing,mignone2008model}.
The available observational Hubble parameter data (OHD) have been
applied in the standard cosmological model
\citep{lin2009observational,stern2010cosmic},  and in some other FRW
models
\citep{samushia2008cosmological,zhang2010constraints,zhai2011constraints}.
Furthermore, the potential of future $H(z)$ observations in
parameter constraint has also been explored \citep{ma2011power}.

SNIa present another widely used observational data in cosmology
research. They give a redshift-distance relationship, the difference
between apparent magnitude and absolute magnitude. SNIa have 580
data points in the latest Union2.1 compilation
\citep{suzuki2012hubble}. Nevertheless, the rich abundance of the
data still can not hide their limitations in determining the
equation of state (EoS)
\citep{garnavich1998supernova,maor2001limitations} and current
deceleration factor $q_0$ \citep{phillipps1983limits} of the
universe. Meanwhile, there have been studies of the limitations in
brane world cosmology \citep{fairbairn2006supernova} and neutrino
radiative lifetimes \citep{falk1978limits}.

Actually, limitations of current observations are mainly attributed
to their measurement relations and model assumptions. For example,
previous CMB, SNIa, weak lensing and BAO, are essentially
\emph{geometric}. In 1962, \citet{sandage1962change} proposed a
promising \emph{dynamical}  survey named redshift drift to directly
probe the dynamics of the expansion. Unlike previous observations,
the redshift drift measures the secular variation of expansion rate
into a deeper redshift of $z=2-5$. It can provide useful information
about the cosmic expansion history in the ``redshift desert", where
other probes are far behind. Unfortunately, extremely weak
theoretical magnitude indicates that it is difficult to detect. For
example, the redshift drift $\Delta z$ within a 10-year
observational time interval for a source at this redshift coverage
has a magnititude of an  order of only $10^{-9}$. The corresponding
velocity drift $\Delta v$ is also inappreciable as only several
cm/s. Fortunately, \citet{loeb1998direct} developed a possible
scheme from the wavelength shift of quasar (QSO) Ly$\alpha$
absorption lines. A new generation of COsmic Dynamics EXperiment
(CODEX) with a high resolution, extremely stable and ultra high
precision spectrograph is now available, capable of measuring such a
small cosmic signal in the near future
\citep{pasquini2005codex,pasquini2006codex}. Based on the power of
CODEX, some groups generate velocity drift by Monte Carlo simulation
with the assumption of standard cosmological model
\citep{liske2008cosmic,liske2008elt,liske2009espresso}. These
simulations investigate the constraints in holographic dark energy
\citep{zhang2007exploring}, modified gravity models
\citep{jain2010constraints}, new agegraphic and Ricci dark energy
models \citep{zhang2010sandage} are investigated. They found that
the velocity drift can provide constraints on models with high
significance. In addition, \citet{balbi2007time} evaluate the
redshift drift from several dark energy models. However, this
evaluation of the power of the redshift drift among current
observations is mostly qualitative, not quantitatively. In this
study, we wish to investigate how many data points or years of
future redshift drift or velocity drift observations could provide
valid constraints on cosmological parameters as good as those from
OHD or SNIa? Furthermore, could future velocity drift data offer
more accurate information about the expansion history?

We attempt to answer these questions via an exploratory, statistical
approach. This paper is organized as follows: In Section
\ref{basic}, we introduce the basic theory of the redshift drift.
Then, we analyze the sensitivity of cosmological parameters to the
Hubble parameter, luminosity distance and the velocity drift.
Section \ref{observation} presents a statistical analysis of these
observations and evaluates the constraint. Section \ref{result}
gives results of constraints in specific evaluation models. In Sec.
\ref{deceleration}, we compare constraints of these observations on
two models of deceleration factor $q(z)$. Finally, Section
\ref{Conclusions} presents  our main results and discussion.

\section{Basic theory}
\label{basic}

\subsection{Redshift drift} \label{redshift drift}

Since the birth of the redshift drift, many observational candidates
like masers and molecular absorptions were put forward, but the most
promising one appears to be the Ly$\alpha$ forest in the spectra of
high-redshift QSOs \citep{pasquini2006codex}. These spectra are not
only distinct from the noise of the peculiar motions relative to the
Hubble flow, but also have a large number of lines in a single
spectrum \citep{pasquini2005codex}. In particular,
\citet{pasquini2005codex} have found that 25 QSOs are presently
known at redshift $z = 2-4$ with a magnitude brighter than 16.5.
Recently, \citet{darling2012toward} shows a set of observational
redshift drift from the precise H{\scriptsize I} 21 cm absorption
line using primarily Green Bank Telescope digital data. These
measurements last 13.5 years for ten objects spanning a redshift of
$z=0.09-0.69$. Table 1 of \citet{darling2012toward} shows that main
redshift drift in this redshift coverage are of order $10^{-8}$
yr$^{-1}$, which is about three orders of magnitude larger than the
theoretical values. The author ascribes this discrepancy to the lack
of knowledge on the peculiar acceleration in absorption line systems
and to the long-term frequency stability of modern radio telescopes.

For an expanding universe, a signal emitted by the source at time
$t_{\mathrm{em}}$  was observed at $t_0$. We represent the source's
redshift through a cosmic scale factor
    \be
    z(t_0) = \frac{a(t_0)}{a(t_{\mathrm{em}})} -1.
    \ee
Over the observer's time interval $\Delta t_0$, the source's
redshift  becomes
    \be
    z(t_0+\Delta t_0) = \frac{a(t_0+\Delta t_0)}{a(t_{\mathrm{em}}+\Delta t_{\mathrm{em}})}-1,
    \ee
where $\Delta t_{\mathrm{em}}$ is the time interval-scale when the
source emits another signal. It should satisfy  $\Delta
t_{\mathrm{em}} = \Delta t_0 / (1+z)$. We represent the observed
redshift variation of the source by
    \be
    \Delta z=\frac{a(t_0+\Delta t_0)}{a(t_{\mathrm{em}}+\Delta t_{\mathrm{em}})}
    - \frac{a(t_0)}{a(t_{\mathrm{em}})}.
    \ee
A further relation can be obtained if we keep the first order
approximation
    \be  \label{change Hubble}
    \Delta z \approx \left[ \frac{\dot{a}(t_0) - \dot{a}(t_{\mathrm{em}})}{a(t_{\mathrm{em}})} \right] \Delta t_0.
    \ee
Clearly, the observable $\Delta z$ is a direct change of the
expansion rate during the evolution of the Universe. In terms of the
Hubble parameter $H(z)=\dot{a}(t_{\mathrm{em}})/a(t_{\mathrm{em}})$,
it can simplify as
   \be  \label{drift def}
   \frac{\Delta z}{\Delta t_0}=(1+z)H_0 - H(z).
   \ee
This is also well known as McVittie equation
\citep{mcvittie1962appendix}. Obviously, cosmological models
associate with the redshift drift just through the Hubble parameter
$H(z)$. Taking a standard cosmological model as an example, we find
that redshift drift at low redshift generally towards to negative
with the dominance of matter density parameter $\Omega_{m}$. This
feature is often used to distinguish dark energy models from LTB
void models at $z<2$ (especially at low redshift)
\citep{yoo2011redshift}. Unfortunately however, the scheduled CODEX
would not be able to measure this drift at low $z$
\citep{liske2008cosmic}. Observationally, it is more common to
detect the spectroscopic velocity drift
    \be  \label{velocity def}
   \frac{\Delta v}{\Delta t_0}= \frac{c}{1+z} \frac{\Delta z}{\Delta t_0},
   \ee
which is in order of several cm s$^{-1}$ yr$^{-1}$. Obviously, the
velocity variation $\Delta v$ can be enhanced with the increase  of
observational time $\Delta t_0$.

For the capability of CODEX, the accuracy of the spectroscopic
velocity drift measurement was estimated by
\citet{pasquini2005codex,whitelock2006scientific} from Monte Carlo
simulations. It can be modelled as
    \begin{equation}\label{velocity error}
\sigma_{\Delta v}=1.4\left( \frac{2350}{\textrm{S/N}} \right)\left(
\frac{30}{N_{\textrm{QSO}}} \right)^{1/2}\left(
\frac{5}{1+z_{\textrm{QSO}}} \right)^{1.8} \mathrm{cm/s},
\end{equation}
where S/N is the signal-to-noise ratio, $N_{\textrm{QSO}}$ and
$z_{\textrm{QSO}}$ are respectively the number and redshift of the
observed QSO. According to currently known QSOs brighter than 16.5
with  $2\la z \la 4$,
\cite{pasquini2005codex,whitelock2006scientific} assumed to observe
either 40 QSOs with S/N ratio of 2000, or 30 QSOs with S/N of 3000,
respectively. In this paper, our investigations are based on the
latter. Unless stated otherwise, the observational time is $\Delta
t_0=10$ years.

\begin{figure}
\centering
\includegraphics[width=0.4\textwidth]{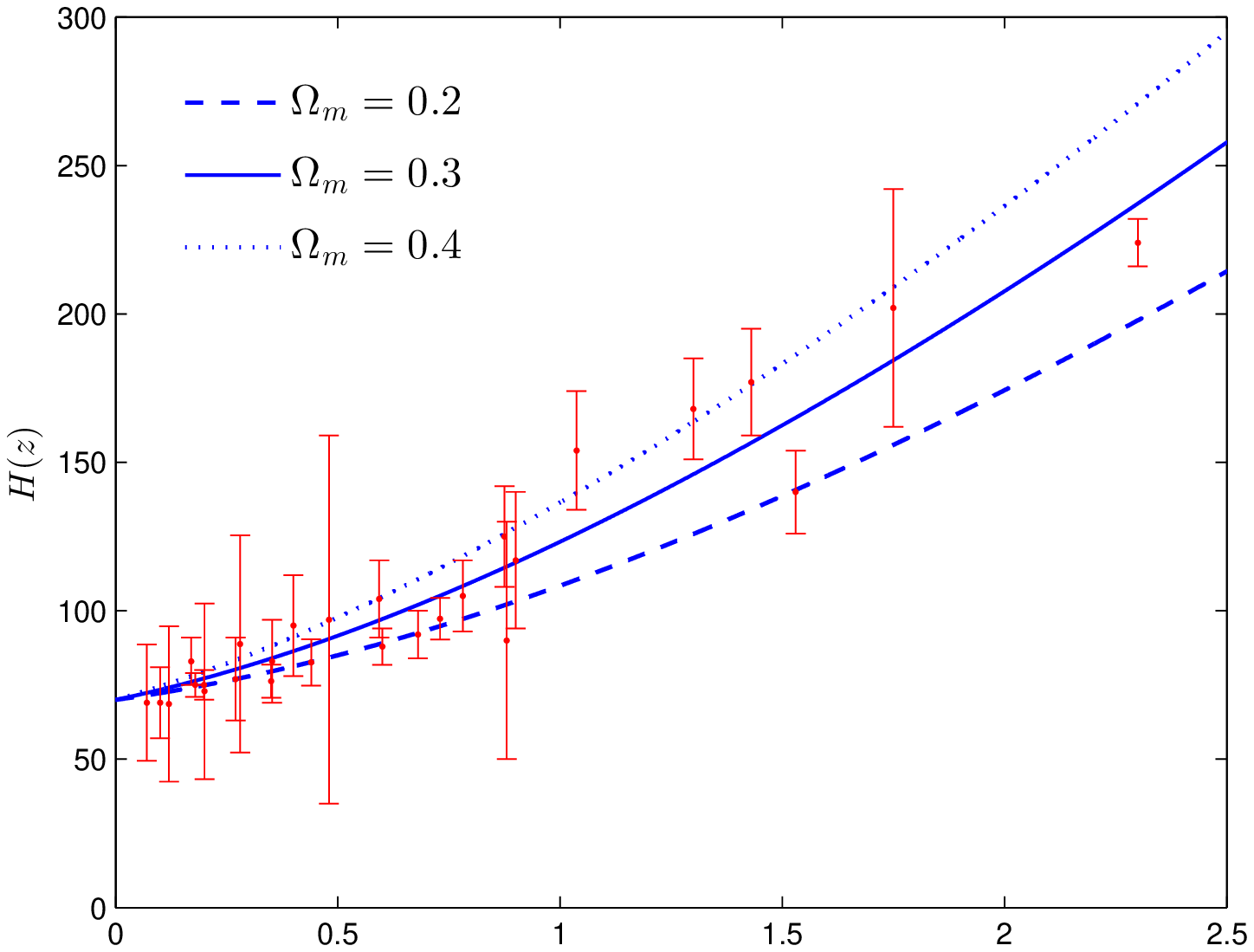}
\includegraphics[width=0.4\textwidth]{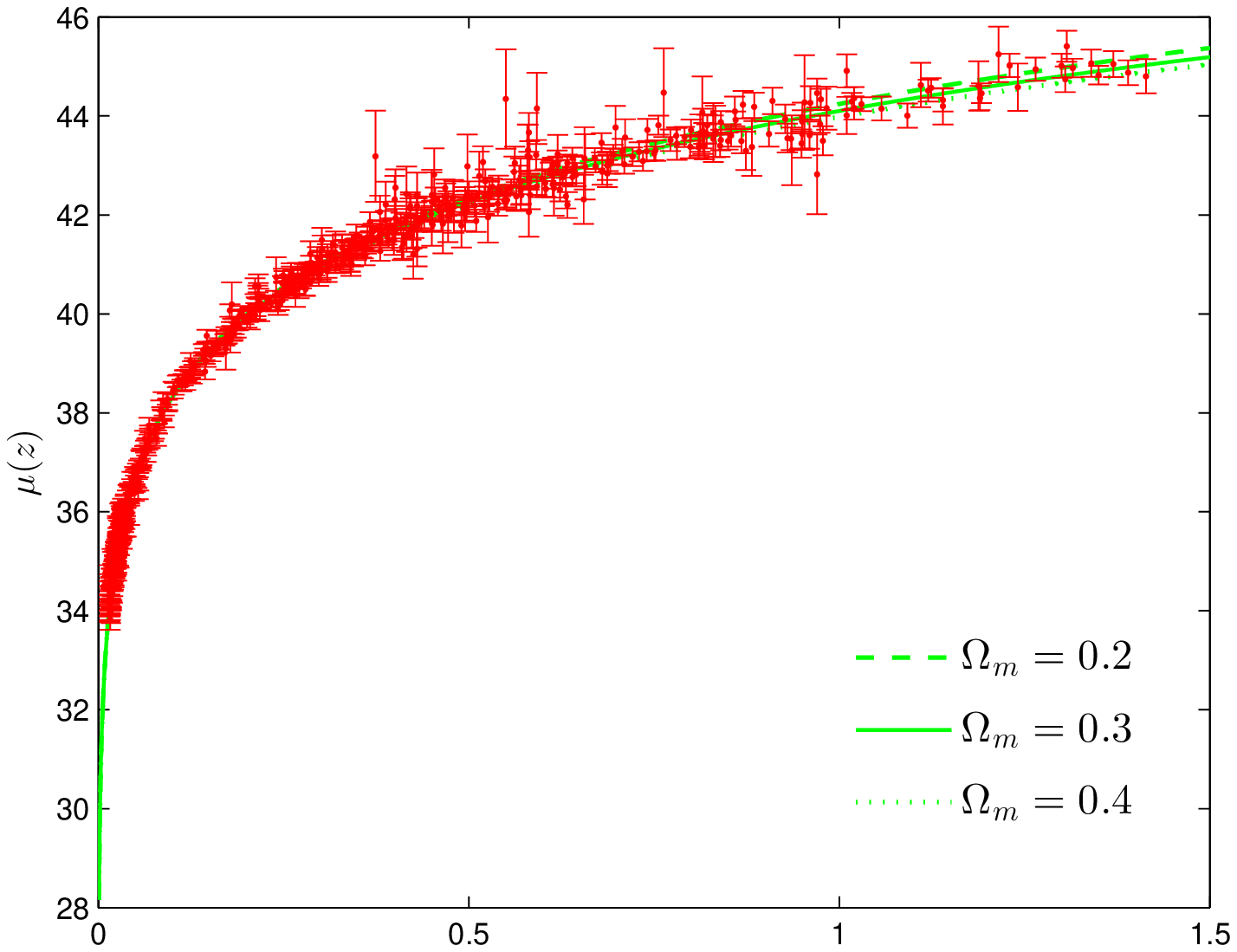}
\includegraphics[width=0.4\textwidth]{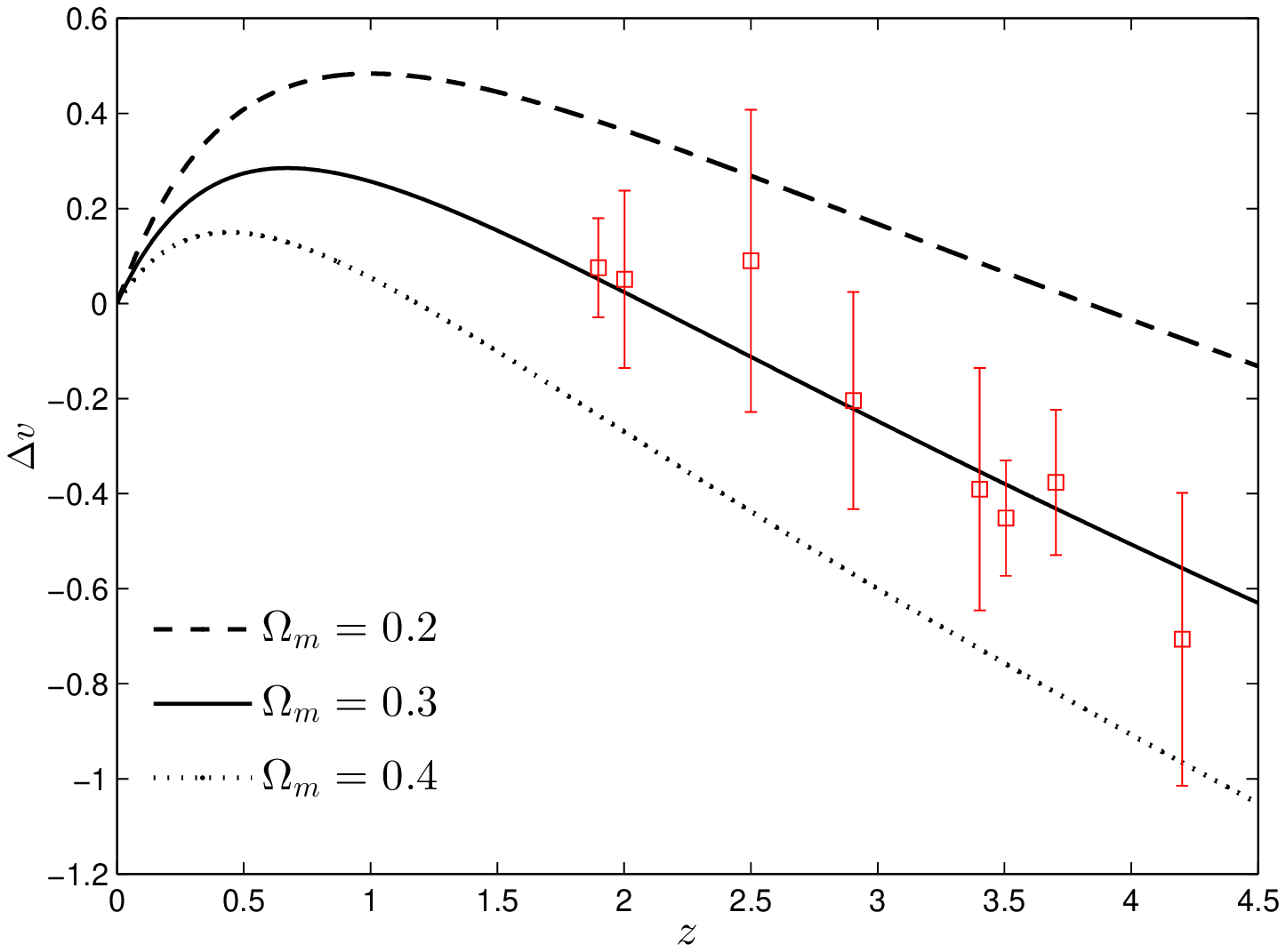}
    \caption{\label{LCDM} Evolution of Hubble parameter, distance modulus and velocity drift
    for the flat $\Lambda$CDM model with different $\Omega_m$.}
\end{figure}

\begin{figure}
\centering
    \includegraphics[width=0.4\textwidth]{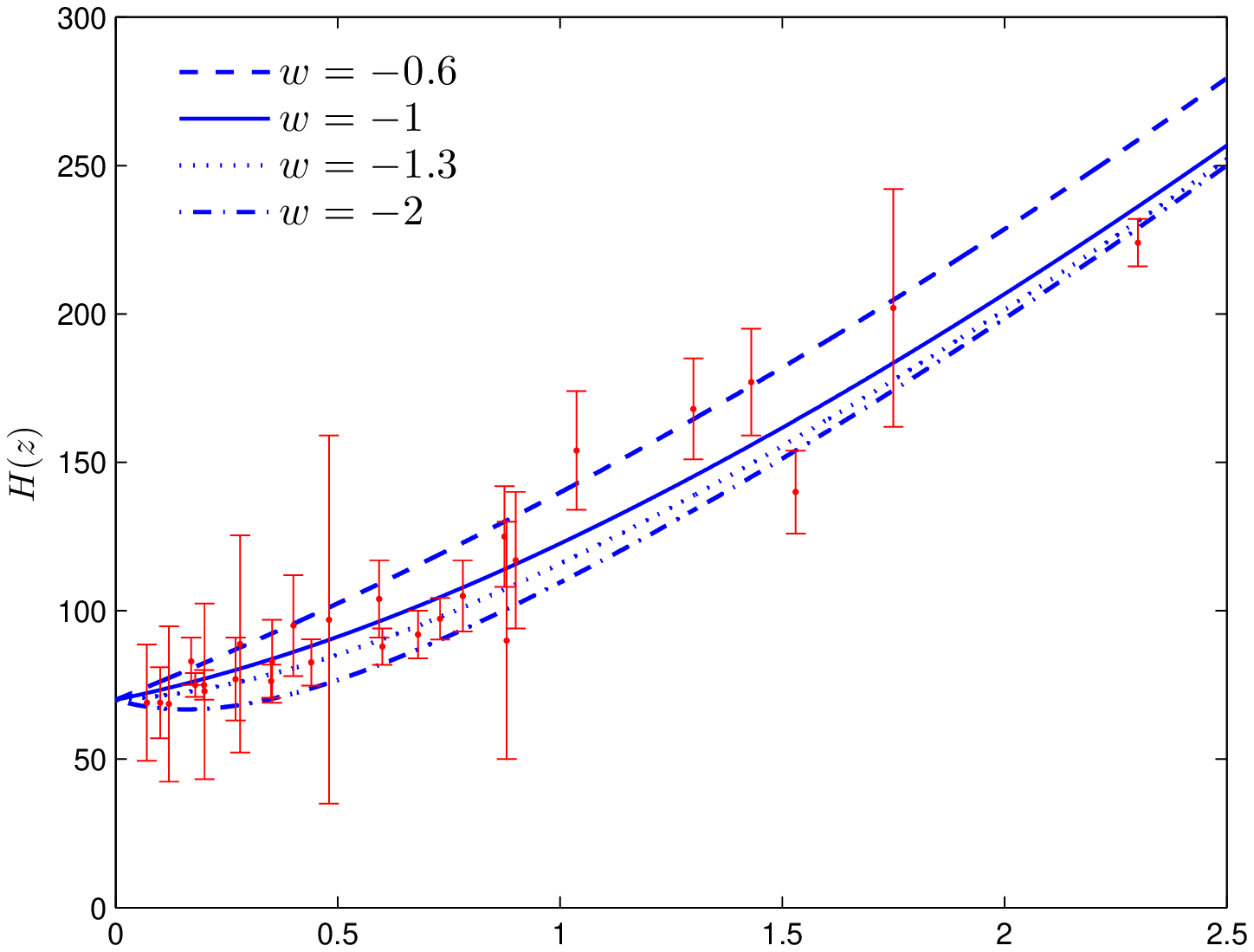}
    \includegraphics[width=0.4\textwidth]{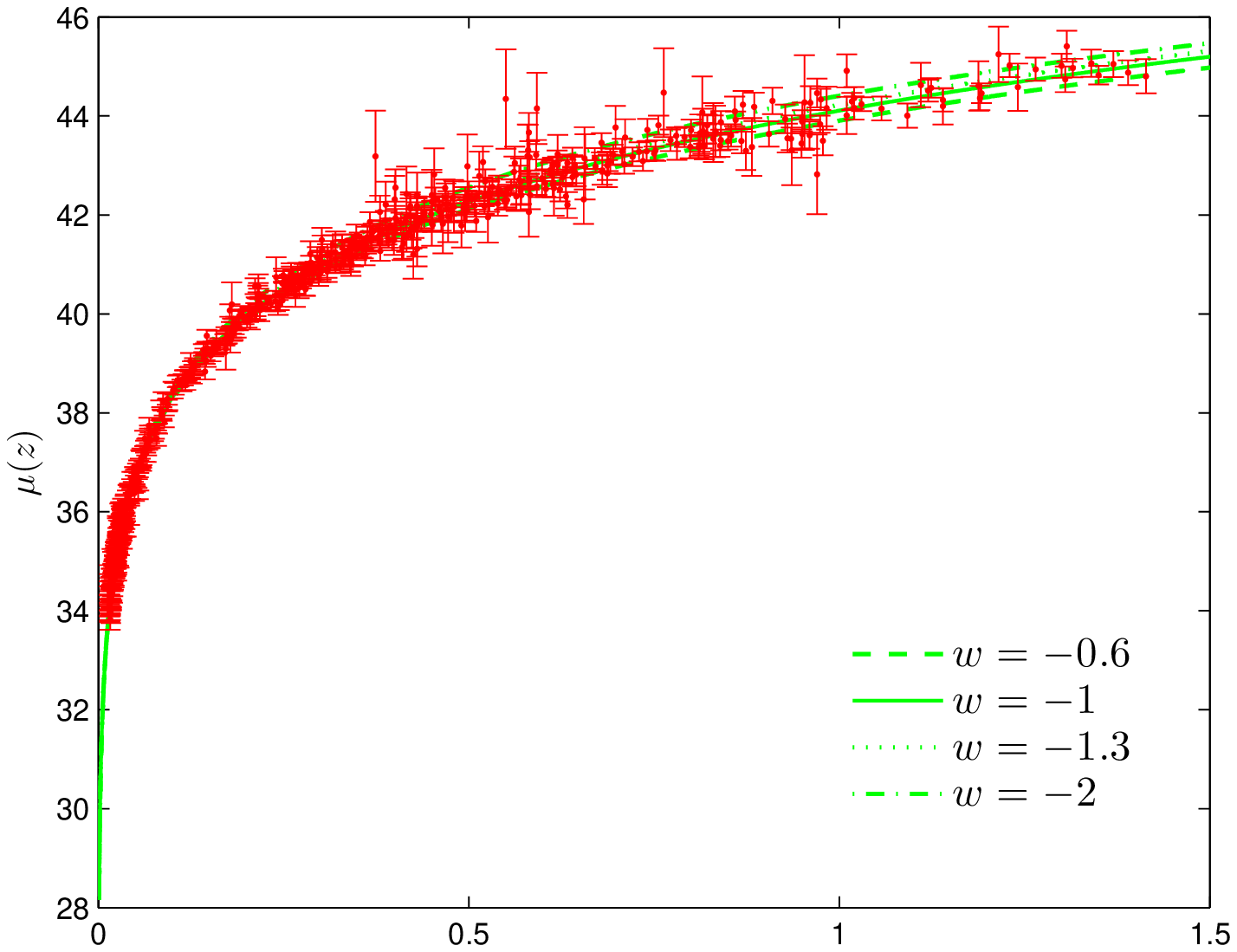}
    \includegraphics[width=0.4\textwidth]{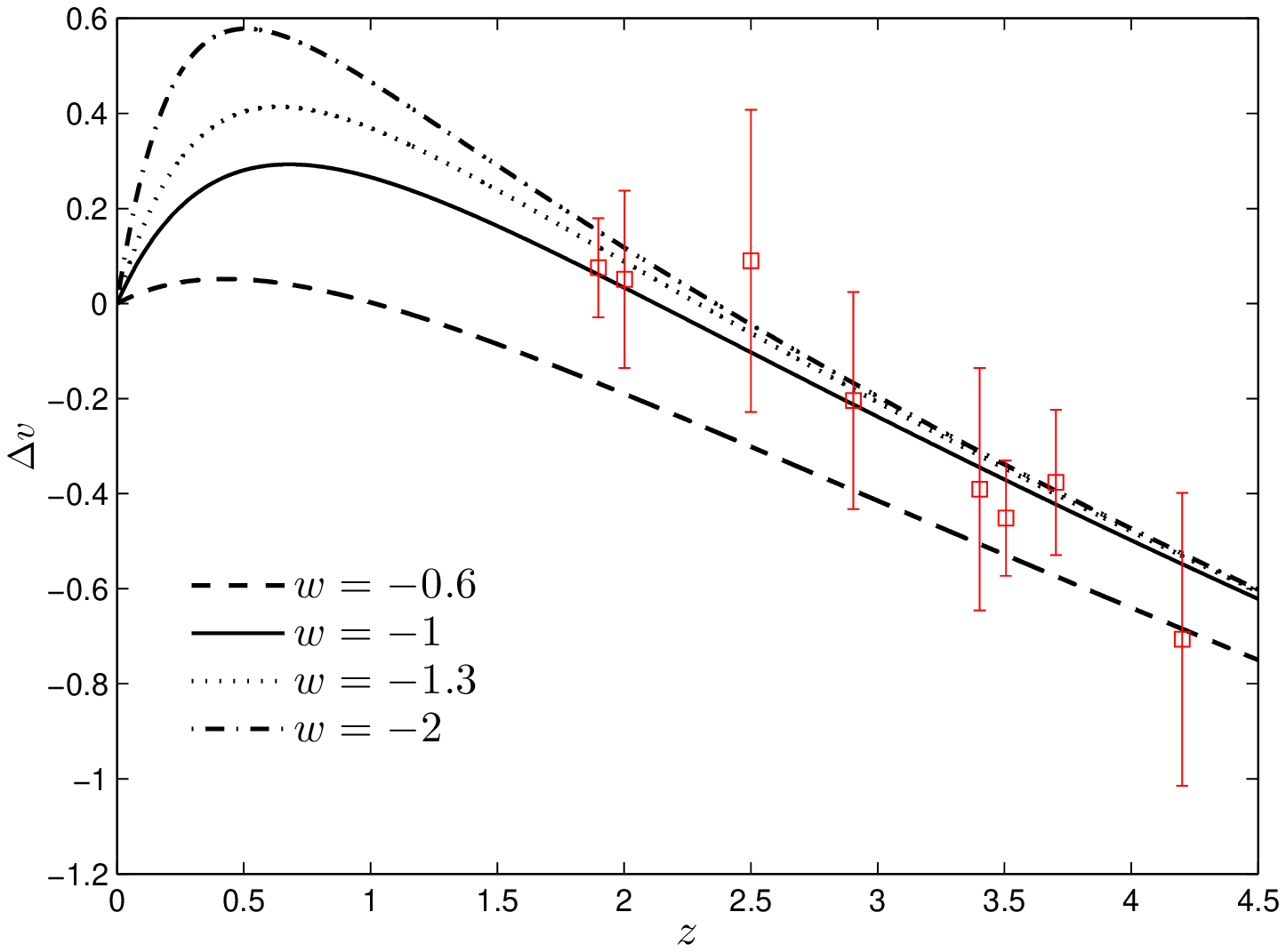}
    \caption{\label{XCDM} Same as Figure \ref{LCDM} but for the flat XCDM model. Dark energy density is fixed at $\Omega_{\Lambda}=0.7$ but for different EoS $w$.}
\end{figure}

\subsection{Sensitivity comparison}
\label{sensitivity}

In fact, most cosmological models can fit well with the
observations. It is difficult to distinguish or rule out some
models. In this paper, we mainly compare the velocity drift with the
OHD and SNIa. For further understanding on some fundamental
parameters, the fiducial cosmological models here are taken as flat
$\Lambda$CDM model and XCDM model.

Changing matter density $\Omega_m$ from 0.2 to 0.4, we plot the
Hubble parameter $H(z)$, distance modulus $\mu(z)$ and velocity
drift $\Delta v$ for the $\Lambda$CDM model in Figure \ref{LCDM}.
Intuitively, we find that the theoretical curves respectively
deviate from each other with different degree. In which, difference
of $\Delta v$ for different $\Omega_m$ becomes remarkable with the
increase of redshift. In order to eliminate the effect of different
units, we take into account the observational data. OHD are the
latest available sample covering redshift $0.07 \leqslant z
\leqslant 2.3$ \citep{farooq2013hubble}. The updated Union2.1
compilation of SNIa  is compiled by \citet{suzuki2012hubble}. Eight
points of $\Delta v$ are simulated by
\citet{liske2008cosmic,liske2008elt,liske2009espresso}  for 20
years. We find that most of current OHD  at $z<1$ can not
distinguish these models, and can not steadily favor a specific
model at $z>1$. While $\mu(z)$ seem to be attracted on the
theoretical curves with little discrimination. Fortunately, only
$\Delta v$ strongly supports the model with $\Omega_m \sim 0.3$.
Different sensitivities show that $\Delta v$ is a possible effective
tool to discriminate $\Lambda$CDM model or accurately determine the
matter density, which agree with the previous investigation
\citep{zhang2007exploring,jain2010constraints}.

In Figure \ref{XCDM}, we plot these three parameters for the flat
XCDM model with fixed $\Omega_{\Lambda}=0.7$. In order to emphasize
the influence of equation of state (EoS) on these parameters, we
change EoS $w$ from -0.6 to -2. Same as above model, $\mu(z)$ in
this case is still insensitive to parameter $w$. Unlike above case,
however, $H(z)$ and $\Delta v$ both are not sensitive to $w$ as long
as $w<-0.6$. However, as estimated from Equation (\ref{velocity
def}), signals can be linearly enhanced with the increase of
observational interval time $\Delta t_0$. Therefore, we possibly
investigate $\Delta v$ for several different years in following
analysis. Moreover, different magnitudes of $\Delta v$ at $z\approx
0.5$ indicate that velocity drift at low redshift may be an
effective scheme to distinguish different dark energy candidates.

\section{Observational constraints}
\label{observation}

In order to explore the answer to questions raised in Section
\ref{introduction}, we respectively introduce constraint methods.
OHD and SNIa  are available to determine parameters by the
$\chi^{2}$ statistics. When observation is absent or not enough, we
can forecast constraint by the Fisher matrix. Constraint from
velocity drift is just finished using this approach.

\subsection{Hubble parameter}
\label{OHD data}

As introduced above, OHD can be measured through the differential
age of passively evolving galaxies and the BAO peaks. We use the
latest available data listed in Table 1 of \citet{farooq2013hubble}.
Parameters can be estimated by minimizing
\begin{equation} \label{OHDchi2}
  \chi^{2}_{\textrm{OHD}}(H_{0}, z, \textbf{p}) = \sum_{i} \frac{[H_0 E(z_i) - H^{obs}
  (z_i)]^2}{\sigma_{i}^{2}},
\end{equation}
where $\textbf{p}$ stands for the parameters vector of each dark
energy model embedded in expansion rate parameter $E(z_i)$. In order
to terminate disturbance of the ``nuisance" parameter, Hubble
constant $H_0$ is integrated as a prior according to the
\citet{collaboration2013planck} suggestion, $H_{0} = 67.3 \pm 1.2$
km s$^{-1}$Mpc$^{-1}$.

\subsection{Luminosity distance}
\label{SN data}

SNIa is famous for its rich abundance of data. The latest Union2.1
compilation \citep{suzuki2012hubble}  accommodates 580 samples. The
future SuperNova Acceleration Probe
(SNAP)\footnote[1]{\url{http://snap.lbl.gov}}  mission is said to be
able to gather high-signal-to-noise calibrated light-curves and
spectra for over 2000 SNIa per year at $0<z<1.7$
\citep{aldering2002overview}. They are usually presented in the
shape of distance modulus, the difference between the apparent
magnitude $m$ and the absolute magnitude $M$
    \be   \label{mu}
    \mu_{\textrm{th}}(z)= m - M = 5 \textrm{log}_{10}D_L(z)+\mu_0,
    \ee
where $\mu_0=42.38-5 \textrm{log}_{10} h$, and $h$ is the Hubble
constant $H_0$ in units of 100 km s$^{-1}$Mpc$^{-1}$. The
corresponding luminosity distance function $D_L(z)$  can be
expressed as
\begin{equation}
    \label{DL}
    D_L(z) =  \frac{1 + z}{\sqrt{\left|\Omega_k \right|}}
    \operatorname{sinn} \left[\sqrt{\left|\Omega_k \right|} \int^z_0 \frac{\mathrm{d}
    z'}{E(z'; \textbf{p})} \right],
\end{equation}
where the $\operatorname{sinn}$ function therein is a shorthand for
the definition
\begin{eqnarray}
    \operatorname{sinn}(x) =
    \begin{cases}
    \sinh x, & \Omega_k > 0, \\
    x, & \Omega_k = 0, \\
    \sin x, & \Omega_k < 0.
   \end{cases}
\end{eqnarray}
Parameters in the expansion rate $E(z'; \textbf{p})$ including the
annoying parameter $h$ commonly determined by the Equation
(\ref{OHDchi2}) but replacing Hubble parameter as distance modulus.
However, an alternative way can marginalize over the ``nuisance"
parameter $\mu_0$
\citep{pietro2003future,nesseris2005comparison,perivolaropoulos2005constraints}.
The rest parameters without $h$ can be estimated  by minimizing
    \be \label{SNchi2}
    \chi^{2}_{\textrm{SN}}(z,\textbf{p})= A-\frac{B^2}{C},
    \ee
where
     \bea
     A(\textbf{p}) &=&  \sum_{i}\frac{[\mu_{\textrm{obs}}(z) - \mu_{\textrm{th}}(z; \mu_0=0,
     \textbf{p})]^2}{\sigma_{i}^{2}(z)},   \nonumber\\
     B(\textbf{p}) &=&  \sum_{i}\frac{\mu_{\textrm{obs}}(z) - \mu_{\textrm{th}}(z; \mu_0=0,
     \textbf{p})}{\sigma_{i}^{2}(z)},   \nonumber\\
     C &=&  \sum_{i} \frac{1}{\sigma_{i}^{2}(z)}.
     \eea
It is equivalent with the general form like Equation
(\ref{OHDchi2}). However, difference from the
$\chi^{2}_{\textrm{OHD}}$ statistics is that $H_0$ in this operation
is marginalized over by Gaussian integration over ($-\infty,
\infty$) without any prior. This program has been used in the
reconstruction of dark energy \citep{wei2007reconstruction},
parameter constraint \citep{wei2010observational},  reconstruction
of the energy condition history \citep{wu2012reconstructing} etc.

\subsection{Velocity drift}
\label{velocity}

\begin{figure*}
\begin{center}
\includegraphics[width=7.5cm,height=6cm]{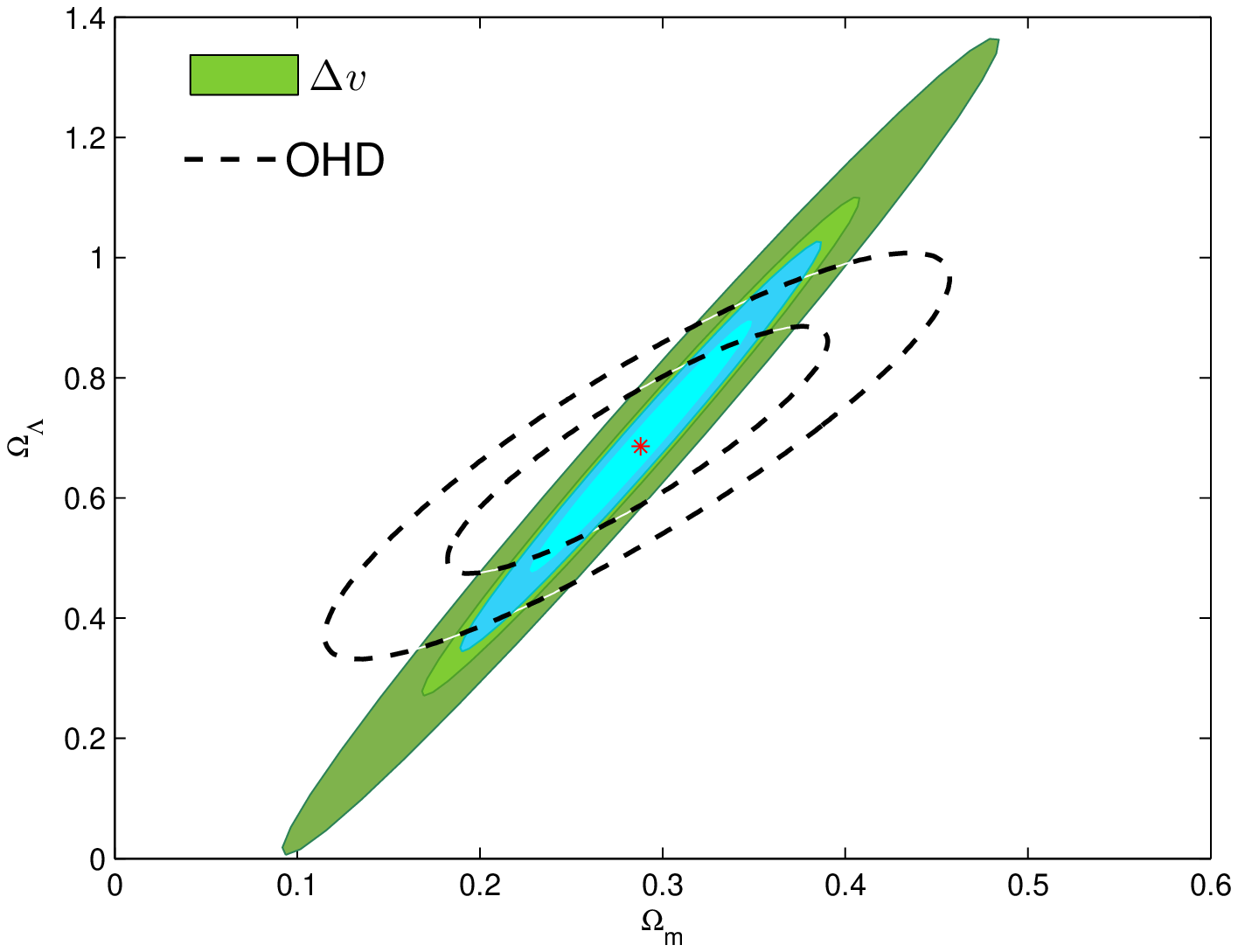}
\includegraphics[width=7.5cm,height=6cm]{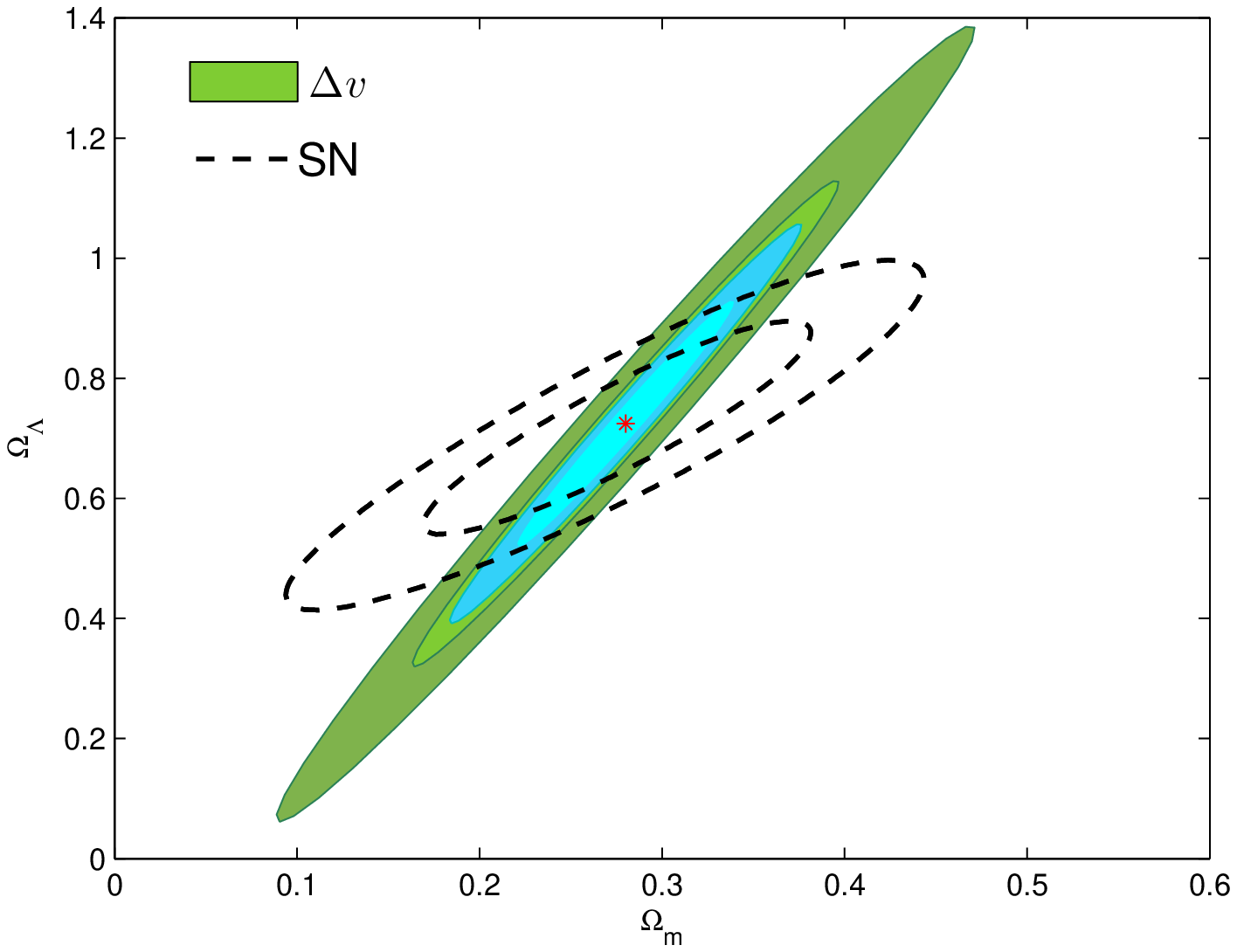}
    \caption{\label{comparisonOHDLCDM} Constraints on the parameters ($\Omega_m$,  $\Omega_{\Lambda}$) for the $\Lambda$CDM model.
    The contours show the 68\% and 95\% C.L., respectively. Dashed curves are contours from available observations.
    Green shaded regions and blue shaded regions are contour constraints of $\Delta v$ for 10 years and 20 years observational
    time, respectively.}
\end{center}
\end{figure*}

Fisher information matrix
\citep{jungman1996cosmological,vogeley1996eigenmode,tegmark1997karhunen,tegmark1997measuring}
could help velocity drift to provide estimation on parameters. This
forecast is a second-order approximation to the likelihood, and has
become an important strategy on parameter constraints in recent
years. Its normal form for velocity drift is
    \be  \label{fisher}
    F_{ij}=\sum \frac{1}{\sigma_{\Delta v}^2} \frac{\partial \Delta v}{\partial
    \theta_i} \frac{\partial \Delta v}{\partial \theta_j},
    \ee
where $\sigma_{\Delta v}$ are errors of $\Delta v$ which has been
estimated from Equation (\ref{velocity error}), $\theta_i$ denotes
the $i$th parameter. For comparison, parameters of Equation
(\ref{fisher}) in the fiducial model are respectively taken to be
the best-fit ones from OHD and SNIa. Therefore, the Fisher matrix
elements, in practice, can be estimated. Eventually, we can compare
the constraint power of $\Delta v$ with OHD and SNIa, respectively.
With the Fisher matrix, we can estimate the uncertainty of parameter
$\theta_i$ through its inverse
    \be
    \Delta \theta_i \geqslant \sqrt{(F^{-1})_{ii}}.
    \ee
The sign $\geqslant$ results from the Cramer-Rao theorem which
states that any unbiased estimator for the parameters is no better
than that from $F^{-1}$. However, on many occasions we need to
produce a Fisher matrix in a smaller parameter space. Similar to
above approach on OHD and SNIa, it can be finished by adding a prior
or marginalizing over the undesired ``nuisance" parameter.  We use
the standard technique issued by the Dark Energy Task Force (DETF)
in XIII. Technical Appendix \citep{albrecht2006report}.

\textbf{Prior}.--- Based on the DETF, we can adopt a Gaussian prior
with error $\sigma$ to the corresponding parameter by adding a new
Fisher matrix $\textbf{F}^{P}$ with a single non-zero diagonal
element $1/\sigma ^2$
    \be
    \textbf{F} \rightarrow  \textbf{F} + \textbf{F}^{P}.
    \ee
For example, $H_0$ should be disposed as a prior when we compare
$\Delta v$ with OHD. In this case, we can simply add $1/\sigma_H ^2$
to $F_{jj}$ where $H_0$ locates.

\textbf{Marginalization}.--- When our attention does not focus on
some nuisance parameters without any additional prior, we can
directly marginalize over them. In the $\chi^2$ statistics,
marginalization is usually defined by integrating the probabilities
on specific nuisance parameter $\theta_i$
    \be   \label{integral}
     \bar{\chi}^2= -2 \textrm{ln} \left( \int^{+\infty}_{-\infty}
e^{-\chi^2/2} d \theta_i \right).
    \ee
The Equation (\ref{SNchi2}) is just accomplished by this method to
marginalize over parameter $H_0$ in normal  $\chi^2$ statistics.
However, in DETF there is a simple way to do this for the Fisher
matrix: Invert \textbf{F}, remove the rows and columns that are
being marginalized over, and then invert the result to obtain the
reduced Fisher matrix \citep{albrecht2006report}.

\subsection{Figure of merit}
\label{fom}

Figure of merit (FoM) is an useful approach to quantitatively
evaluate the constraining power of cosmological data. It has been
used to evaluate constraint power of some simulated data on the dark
energy EoS
\citep{albrecht2006report,albrecht2009findings,acquaviva2010falsify,wang2010designing,ma2011power},
or to choose which available data combination is optimal
\citep{wang2008figure,mortonson2010figures}. Nevertheless, we note
that some versions of FoM are proposed, such as the DETF to
constrain ($w_0$, $w_a$) \citep{albrecht2006report}, a simple one
from constraint area \citep{mortonson2010figures} or a generalized
one from determinant of the covariance matrix
\citep{wang2008figure}. Recently, \citet{sendra2012supernova}
investigate their relations. Moreover, \citet{linder2006biased}
tests the concept with significant accuracy. In fact, these concepts
in essence are unanimous, namely reward a tighter constraint but
punishing a looser one. In this paper, we use a statistical
definition \citep{mortonson2010figures} similar to above ones
    \be
    \textrm{FoM} \approx \frac{6.17 \pi}{A_{95}},
    \ee
where $A_{95}$ is the enclosed area of constrained parameters space
at 95\% confidence. For this version, we will marginalize over some
undesired parameters as insurance for 2D constraint. Obviously, the
smaller the area is, the larger the FoM becomes. One remark is that
our definition of FoM is purely statistical rather than physical.

\begin{figure}
\begin{center}
\includegraphics[width=7.5cm,height=6cm]{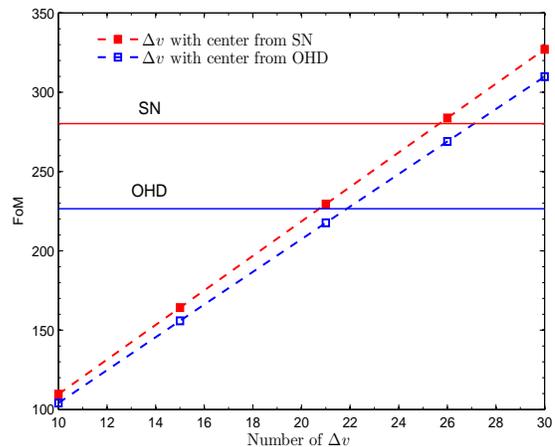}
    \caption{\label{fomLCDM} FoM of each observation for the $\Lambda$CDM model. Horizontal
       lines across the figure mark the FoM of OHD and SNIa. Dashed lines are FoMs of $\Delta v$ from the Fisher matrix
forecast with 10 years. Line marked with face is estimated at
central values constrained by SNIa,
       while line without face is from that of OHD.}
\end{center}
\end{figure}

\section{Results for the evaluation models}
\label{result}

In order to comprehend cosmological density parameters and EoS, we
respectively evaluate above observational data for a standard
non-flat $\Lambda$CDM model and XCDM model.

\begin{figure*}
\begin{center}
\includegraphics[width=0.45\textwidth]{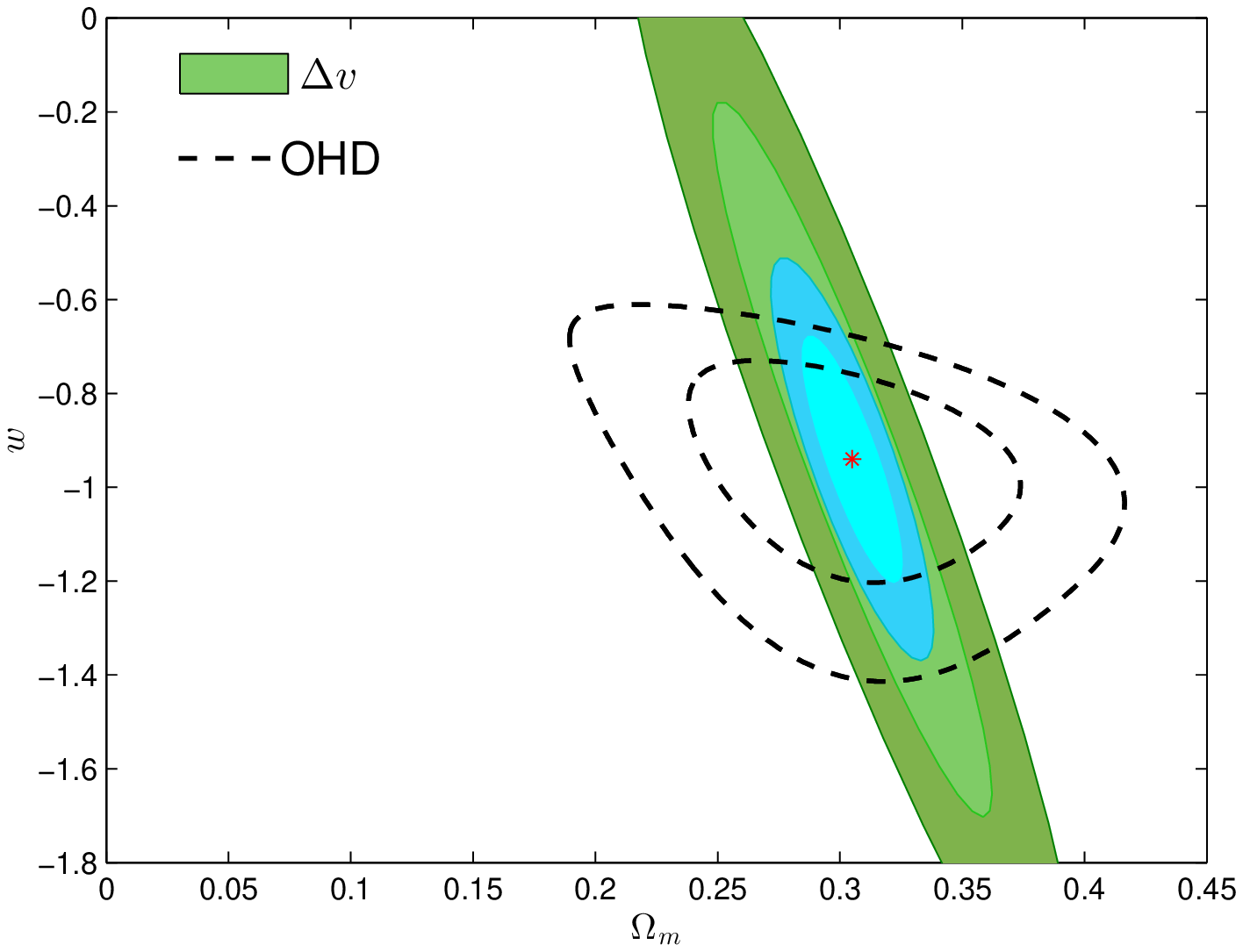}
\includegraphics[width=0.45\textwidth]{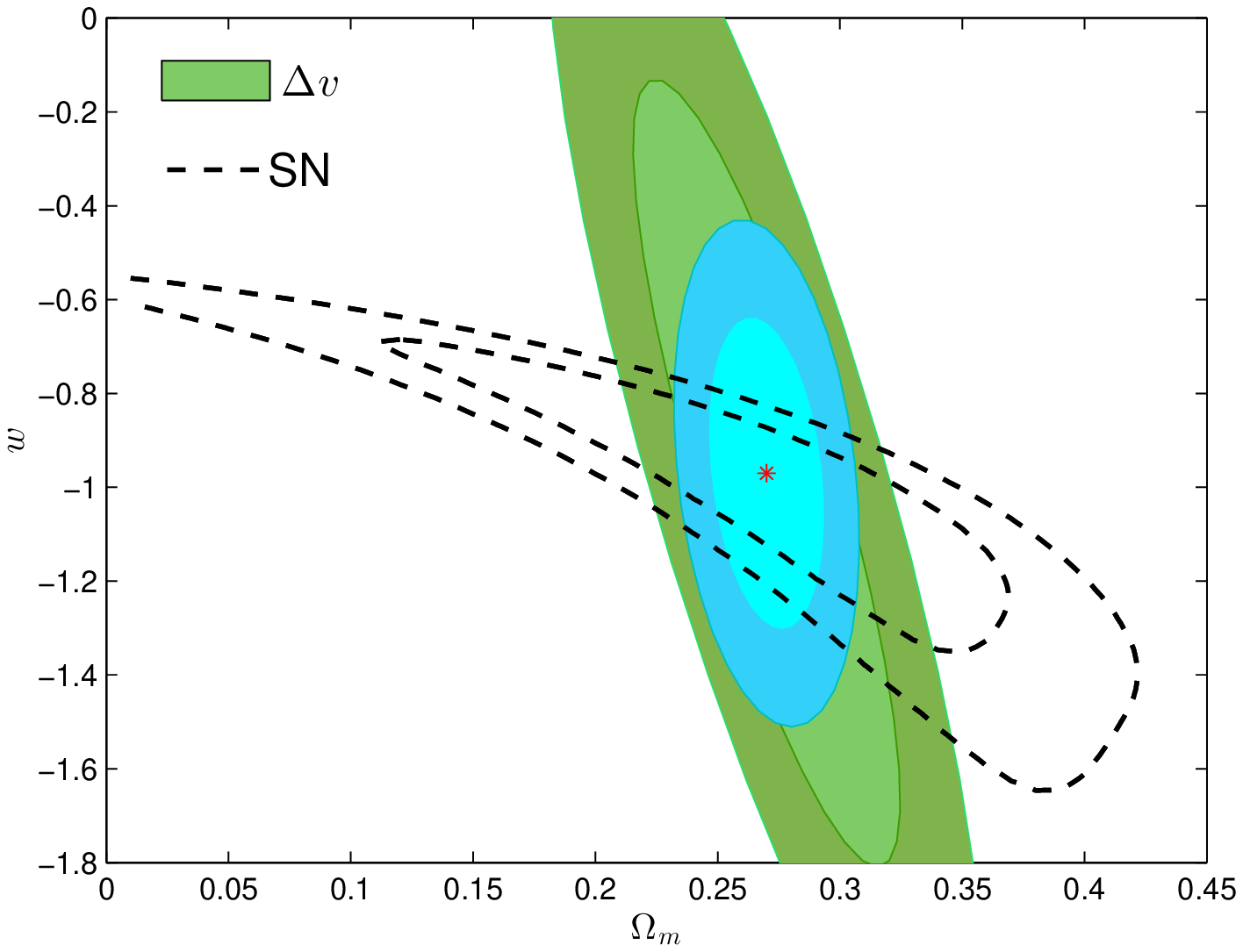}
    \caption{\label{comparisonOHDXCDM} Joint two-dimensional marginalized constraint on parameters ($\Omega_m$,  $w$) of the XCDM model.
    Dashed lines are contours from observational data.  Green shaded regions and blue shaded regions are contour constraints of $\Delta v$ for 10 years and 30 years observational
    time, respectively.}
\end{center}
\end{figure*}
\subsection{$\Lambda$CDM model}
\label{LCDM model}

In the $\Lambda$CDM model, cosmological constant is believed to be
the impetus of the accelerating expansion. The Hubble parameter in
such model is given by
     \be   \label{LCDM hubble}
   H(z)=H_0\sqrt{\Omega_m (1+z)^3  + \Omega_k (1+z)^2 + \Omega_{\Lambda}}.
     \ee
Using the normalization condition on space curvature
$\Omega_k=1-\Omega_m-\Omega_{\Lambda}$, the free parameters are
($\Omega_m$, $\Omega_{\Lambda}$, $H_0$). According to introduction
in Section \ref{observation}, OHD give constraints
$\Omega_m=0.2880^{+0.0670}_{-0.0703}$ and
$\Omega_{\Lambda}=0.6860^{+0.1330}_{-0.1385}$. Using the Fisher
matrix of $\Delta v$ in this case, we present its constraint in
Figure \ref{comparisonOHDLCDM} for different years. We find that 30
QSOs with 10 years present uncertainties $\Delta \Omega_m =0.0790$
and $\Delta \Omega_{\Lambda} =0.2738$, a similar estimation as OHD
on matter density but a larger one on dark energy. For more years,
such as 20 years, uncertainties can be condensed as 0.0398 and
0.1374, respectively. Obviously, uncertainty of matter density is
much tighter than that of OHD. This is predictable because of high
sensitivity of $\Omega_m$ on velocity drift. By the $\chi^2$
statistics, SNIa  presents a similar matter density
$\Omega_m=0.2800^{+0.0677}_{-0.0724}$ and a little bigger dark
energy density $\Omega_{\Lambda}=0.7245^{+0.1131}_{-0.1214}$. It can
be seen that, power of 28 OHD on standard cosmological model is
enough to match that of 580 SNIa. For the $\Delta v$ at this
fiducial model,  bottom panel shows that observation within 10 years
is not enough to determine a better uncertainty estimations on these
parameters than those of SNIa. However, we find that uncertainty
estimation on $\Omega_m$ from $\Delta v$ within 20 years enhance
much more, even its uncertainty at 2$\sigma$ confidence level can be
comparable with that from SNIa at 1$\sigma$ confidence level.
Specifically, our calculation indicates that uncertainty of matter
density would been narrowed for 42\%, i.e., $\Delta
\Omega_m=0.0387$. In a short, ten years of $\Delta v$ are enough to
determine a similar estimation as current observations on the matter
density, but twenty years are needed to determine the dark energy
density.

Besides the uncertainty estimation, purely statistic FoM  also could
provide an evaluation on observations. Inverse of the area of 95\%
confidence region in parameters ($\Omega_m$,$\Omega_{\Lambda}$)
panel multiplying by a positive constant shows that FoMs of OHD and
SNIa are 226.54 and 280.22, respectively. According to the
constraints from OHD and SNIa,  FoM of $\Delta v$ can be
respectively estimated. Therefore, it is naturally divided into two
groups. Modelled capability of CODEX indicates that it can
accommodate 30 QSOs with S/N of 3000. We investigate different
amounts of data points with 10 years. Our main results are shown in
Figure \ref{fomLCDM}. One may intuitively observe that the FoM of
$\Delta v$  linearly increases with the size of the data set. And
each FoM does not change much for different fiducial models, namely,
different central values. We find that 21  $\Delta v$  lead to an
FoM of 229.48, which could reach the parameter constraint power of
OHD. Comparing with the SNIa,  FoM of 26 $\Delta v$ data is 283.72,
which could serve a similar constraint as SNIa. For 30 QSOs, FoM of
$\Delta v$ has reached 327, much better than those from the
observational data.

\begin{figure}
\begin{center}
\includegraphics[width=0.4\textwidth]{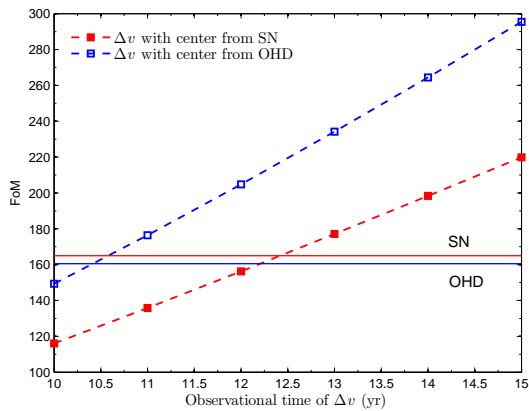}
    \caption{\label{fomXCDM} FoM of each observation for the XCDM model. Marks in this figure are same as Figure \ref{fomLCDM}. However,
    FoMs
    of $\Delta v$ in this case are estimated for different years with 30 QSOs.}
\end{center}
\end{figure}

\begin{figure}
\centering
\includegraphics[width=0.4\textwidth]{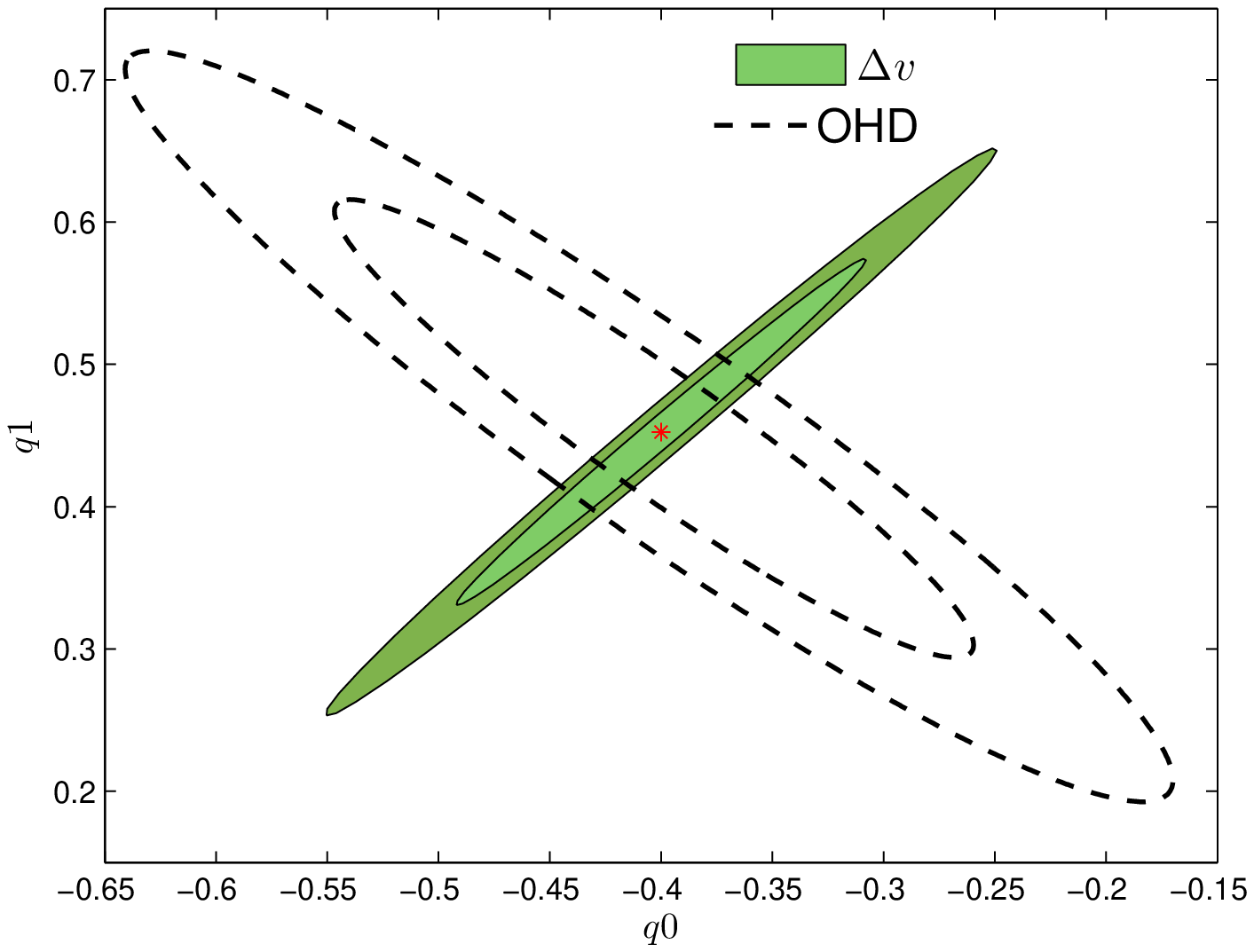}
\includegraphics[width=0.4\textwidth]{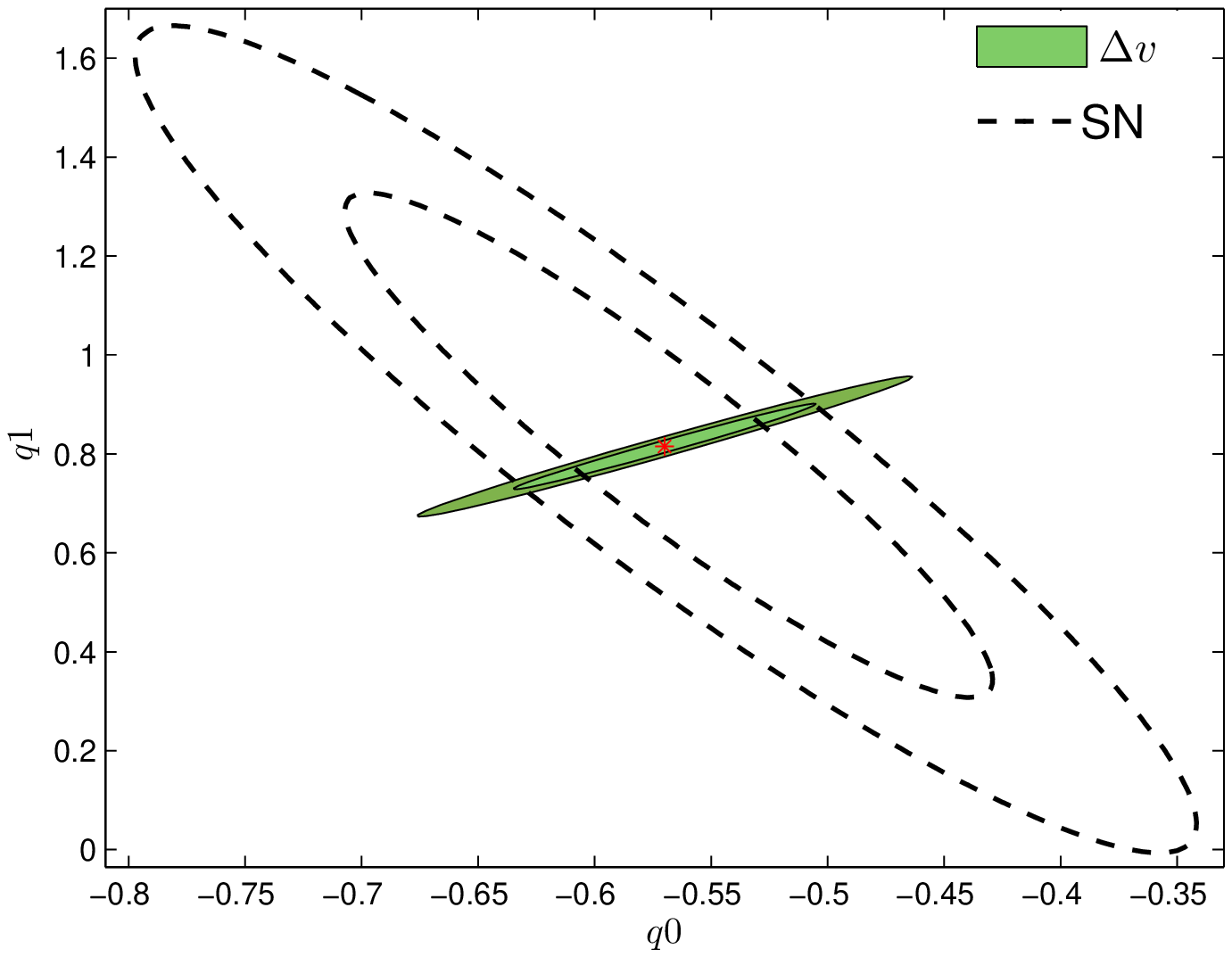}
    \caption{\label{q1comparison} Constraint on parameters ($q_0$, $q_1$) in deceleration factor of model I: $q(z)=q_0 + q_1 z$.}
\end{figure}

\subsection{XCDM model}
\label{XCDM model}

Expansion rate in a non-flat FRW universe with a constant EoS is
given by
   \be   \label{w hubble}
   H(z)=H_0\sqrt{\Omega_m (1+z)^3  + \Omega_k (1+z)^2 + \Omega_{\Lambda}(1+z)^{3(1+w)}}.
   \ee
Difference from the $\Lambda$CDM model is that EoS may be a value
deviation from -1. Besides the cosmological constant with $w= -1$,
dark energy candidates generally can be classified as quiessence
with $-1< w <0 $, quintessence with $-1< w <1$, and phantom with $w
<-1$.

Figure \ref{XCDM} shows that these observations are difficult to
distinguish dark energy candidates. It can be witnessed in this
section. Same operation as above model, OHD give
$w=-0.94^{+0.1373}_{-0.1596}$ slightly off cosmological constant.
After marginalizing over the residual dark energy density parameter,
matter density parameter and $w$ present a closed contour relation
as shown in top panel of Figure \ref{comparisonOHDXCDM}.
Nevertheless,  constraint of $\Delta v$ in this case for 10 years is
not optimistic. It is mainly because of looser uncertainty
estimation on $w$. With increase of observational time, we find that
power of $\Delta v$  on $w$ with 30 years could be comparable with
OHD. From ten to thirty years, uncertainties  of $w$ from $\Delta v$
correspondingly improve from 0.5028 to 0.1729, which reduces the
$\Delta w$ by a factor of three. Meanwhile, $\Delta \Omega_m$ in
this case for velocity drift is 0.0133, which is superior than OHD
three times.

For the SNIa, they present an estimation near the cosmological
constant, $w=-0.97^{+0.1985}_{-0.2250}$. $\Delta v$ in this model
for 10 years does not place a tight constraint on $w$, but a better
constraint on $\Omega_m$. Increasing the observational time for 10
years to 30 years, $\Delta w$ could improve from 0.5536 to 0.2176.
That is, physically, at least thirty years are needed to catch the
constraint power of current observations. Moreover, $\Omega_m$ in
this case is almost constrained with no degeneracy. The $\Delta v$
constraint with 30 years is almost orthogonal to that provided by
SNIa, which also provide a possibility that joint constraints
between them may determine parameters with high significance.

Making the measurement of FoM, OHD and SNIa provide an FoM of 160.5
and 164.9, respectively. Unlike above model, they are largely
identical but with minor differences. Assuming 30 QSOs are
monitored, we could extend the observational time to place tight
constraint. Figure \ref{fomXCDM} shows that at least 12 years are
needed to reach the FoM level of OHD and SNIa.

\begin{figure}
\centering
\includegraphics[width=0.4\textwidth]{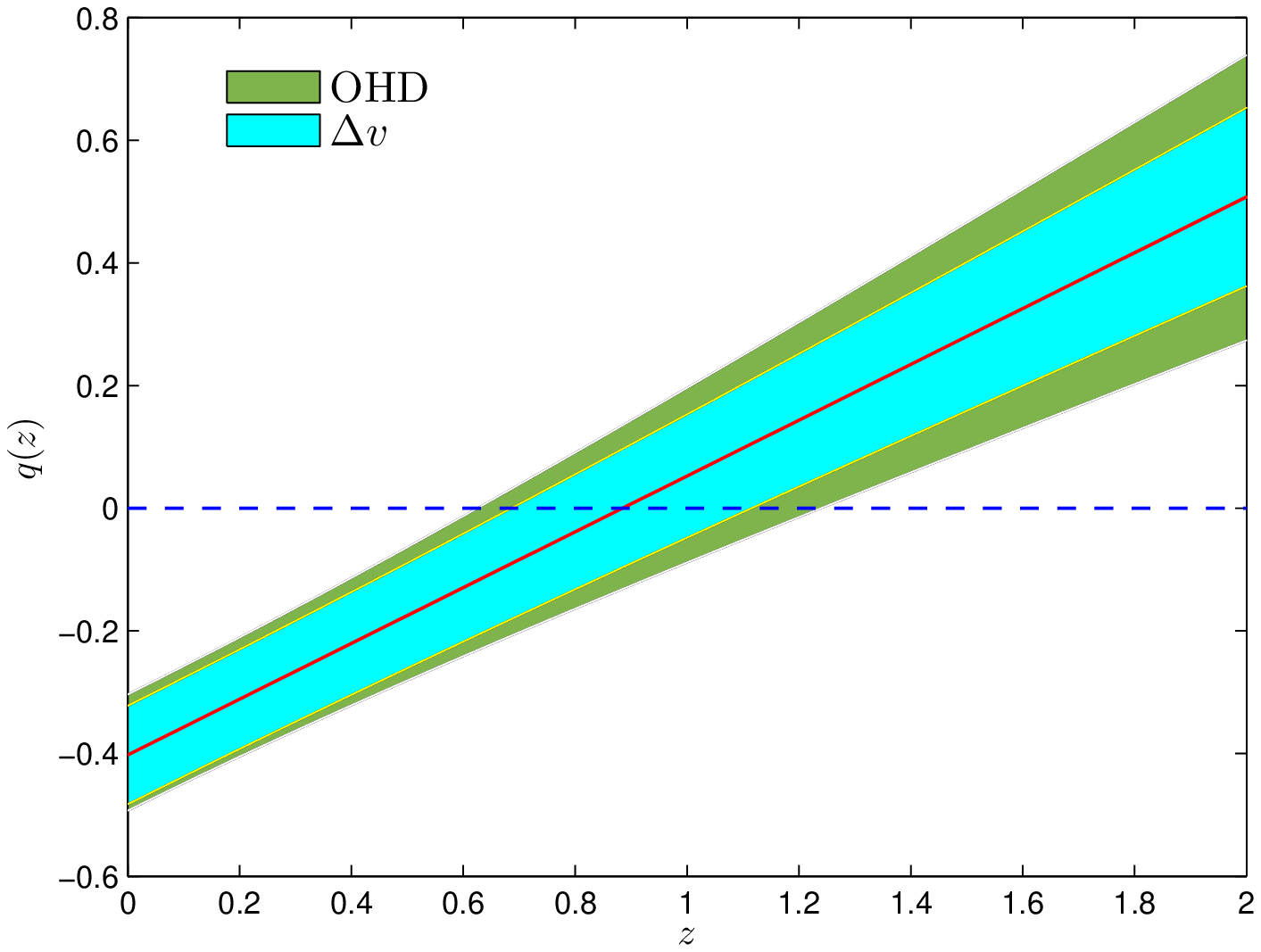}
\includegraphics[width=0.4\textwidth]{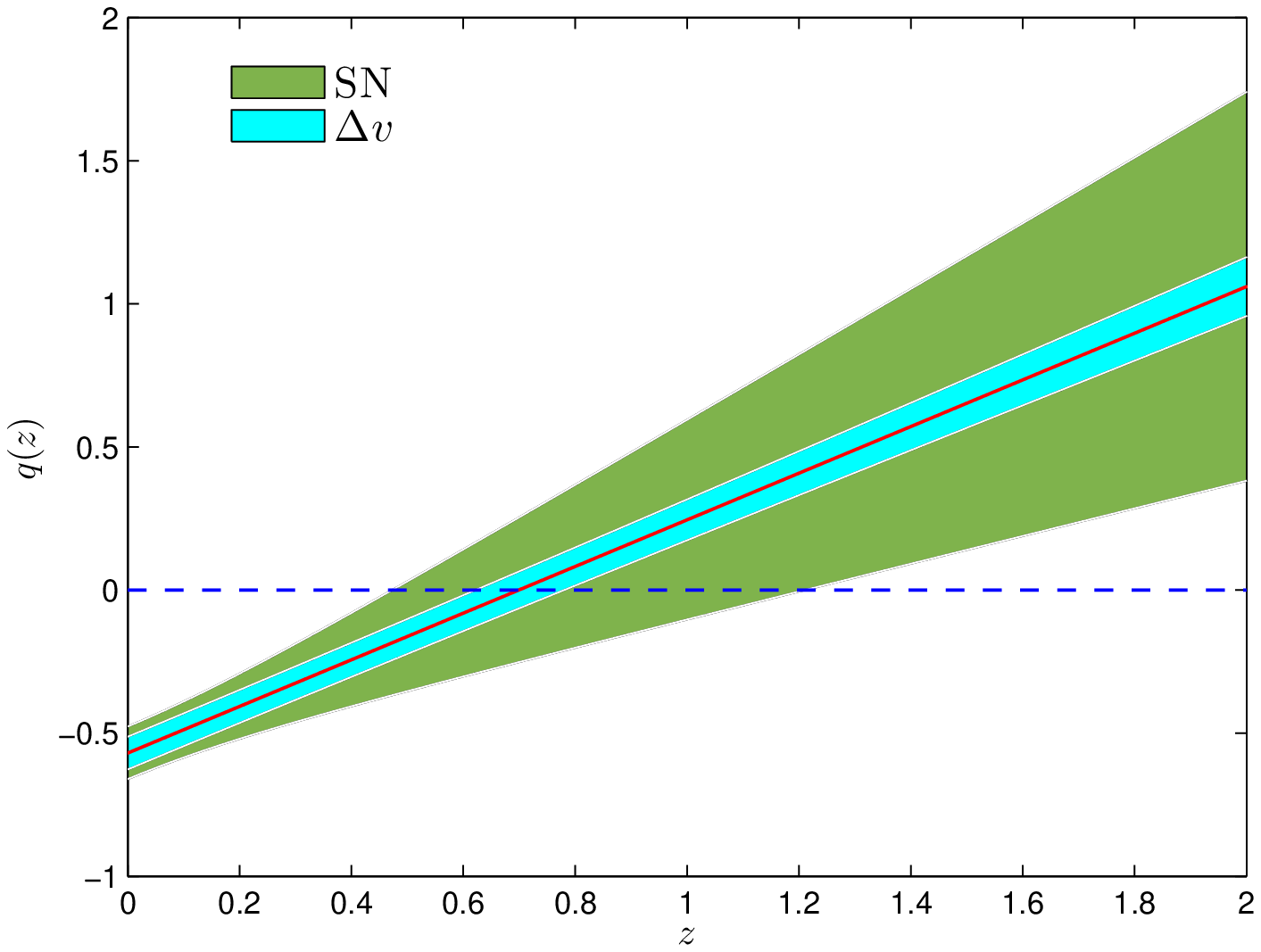}
    \caption{\label{q1reconstruction} Reconstruction of deceleration factor for model I: $q(z)=q_0 + q_1
    z$, using OHD, SNIa and $\Delta v$ with 10 years.}
\end{figure}

\section{Constraint on deceleration factor}
\label{deceleration}

As Equation (\ref{change Hubble}) shown, one difference of the
velocity drift aims at variation of the first derivative of
cosmological scale factor. How well does this dynamical probe offer
more accurate information about the expansion history? Here, we
would like to investigate two parameterized deceleration factor.

Deceleration factor can be defined as
    \be  \label{q definition}
    q(z)=-\frac{\ddot{a}}{aH^2} = \frac{1}{E(z)} \frac{d E(z)}{d z}(1+z) -1.
    \ee
As known from this equation, $q(z)<0$ corresponds to the
accelerating expansion, $q(z)>0$ means a deceleration.  The
transition redshift where the expansion of the universe switched
from deceleration to acceleration is our common focus. Moreover,
previous deceleration leads to $d q/d z>0$. According to the
Equation (\ref{q definition}),  the dimensionless Hubble parameter
can be written as
    \be
    E(z)=\textrm{exp} \left[\int^{z}_{0} [1+q(z')]d \textrm{ln}(1+z') \right].
    \ee
Finally, specific deceleration factor $q(z)$ can be reconstructed
through $E(z)$ in observational variables, such as distance modulus
of Equation (\ref{mu}) and velocity drift of Equation (\ref{velocity
def}). In particular, this reconstruction no longer depends on
cosmological dark energy models. Note that reconstruction from SNIa
is explicitly dependent on one arbitrary constant, namely, the
curvature parameter. Technically, $\Omega_k$ appearing in Equation
(\ref{DL}) is marginalized. Since \citet{riess2004type} raised a
linear $q(z)$, much more parameterizations have been put forward.
Two ordinary models are examined here. Although they have been
investigated by \citet{cunha2008transition,cunha2009kinematic} using
different samples of Supernova Legacy Survey, our further test
mainly emphasizes the power of a new future observation, i.e., the
velocity drift.

\subsection{model I}
\label{model1}

\begin{figure}
\centering
\includegraphics[width=0.4\textwidth]{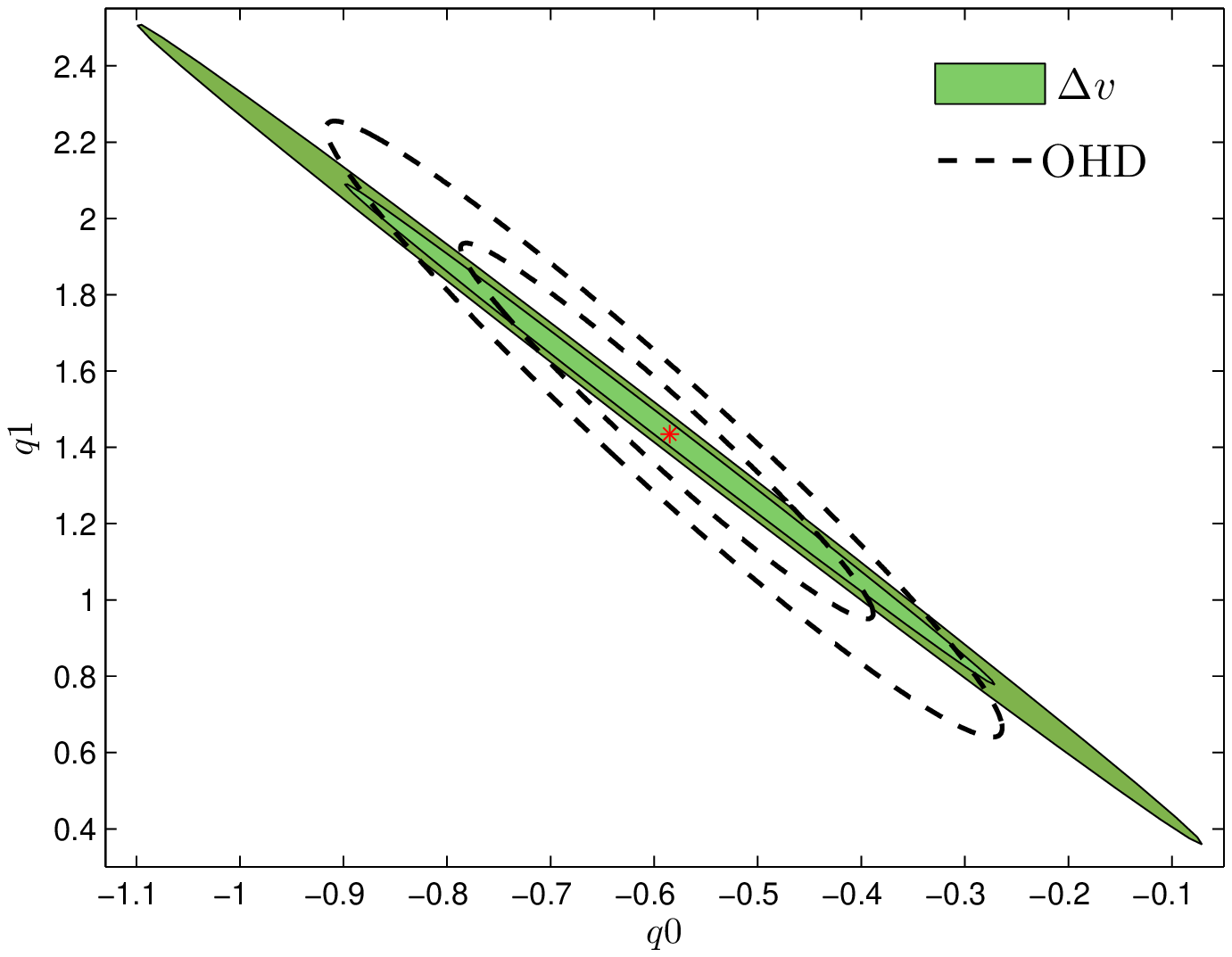}
\includegraphics[width=0.4\textwidth]{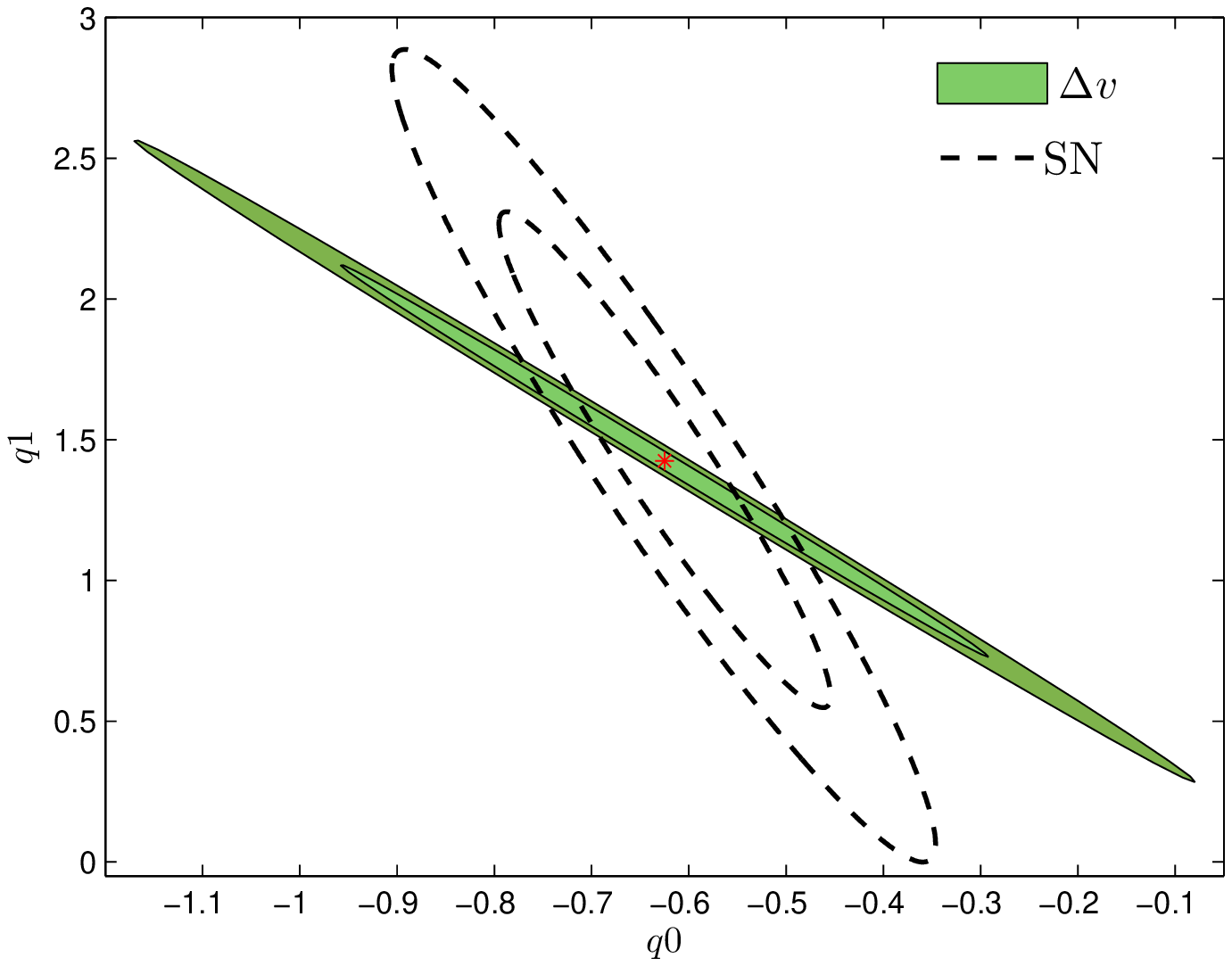}
    \caption{\label{q2comparison}  Constraint on parameters ($q_0$, $q_1$) in deceleration factor of model II: $q(z)=q_0 + q_1 z/(1+z)$.}
\end{figure}

The simplest model for the deceleration parameter is parameterized
by \citet{riess2004type}
    \be  \label{q model1}
    q(z)=q_0 + q_1 z,
    \ee
where $q_0$ is the deceleration factor today, constant $q_1$ is its
change rate. Transition from deceleration to acceleration,
therefore, occurs at redshift $z_t=-q_0/q_1$.

After a prior over $H_0$, OHD provide $q_0=-0.4^{+0.0930}_{-0.0963}$
and $q_1=0.4525 ^{+0.1064}_{-0.1057}$, which indicates a recent
acceleration and previous deceleration. Uncertainties estimated by
$\Delta v$ are $\Delta  q_{0}=0.0802$ and $\Delta  q_{1}=0.0607$,
which are smaller than those of OHD, especially the variation rate.
In Figure \ref{q1comparison}, individual constraints are shown. The
$\Delta v$ constraint is almost orthogonal to that of the OHD.
Therefore, joint constraints may help us more to break the
degeneracy between $q_0$ and $q_1$. Estimations on the parameters
thus can be greatly improved. The deceleration factor in Equation
(\ref{q model1}) is reconstructed by OHD in Figure
\ref{q1reconstruction}, and a transition is found at redshift
$z_t=0.88^{+0.3555}_{-0.2588}$, which is in good agreement with
recent determination of $z_{\textrm{da}}=0.82 \pm 0.08$ based on 11
$H(z)$ measurements between redshifts $0.2\leqslant z \leqslant 2.3
$ \citep{2013A&A...552A..96B}. $\Delta v$ with 10 years gives
$z_t=0.88 ^{+0.2307}_{-0.1988}$, a little tighter than OHD. We
believe that much better estimation can be obtained for more years.

After marginalization over curvature $\Omega_k$, SNIa provide
$q_0=-0.57^{+0.0927}_{-0.0906}$ and $q_1=0.815^{+0.3368}_{-0.3359}$.
We find that uncertainty of $q_0$ between OHD and SNIa are nearly
the same. But the latter presents a higher change rate $q_1$ and
looser uncertainties. This is due to the luminosity distance, an
integral relation of Hubble parameter
\citep{sahni2000case,starobinsky1998determine}
    \be
    H(z)=\left[\frac{d}{dz}\left( \frac{D_L(z)}{1+z} \right) \right]^{-1}.
    \ee
It could smear out many information about the expansion history.
Reconstructed $q(z)$ in bottom panel of Figure
\ref{q1reconstruction} indicates a later transition at
$z_t=0.69^{+0.5018}_{-0.2284}$, which agree with
\citet{cunha2008transition,cunha2009kinematic}. $\Delta v$ in this
case estimates uncertainties $\Delta q_{0}=0.0571$ and $\Delta
q_{1}=0.0427$, which are much tighter than those of SNIa. Slender
outline of reconstruction realizes that the $\Delta v$ is more
powerful than SNIa. A much narrower constraint is therefore
obtained, $z_t=0.69^{+0.0813}_{-0.0774}$.

\subsection{model \textrm{II}}
\label{model2}

\begin{figure}
\centering
\includegraphics[width=0.4\textwidth]{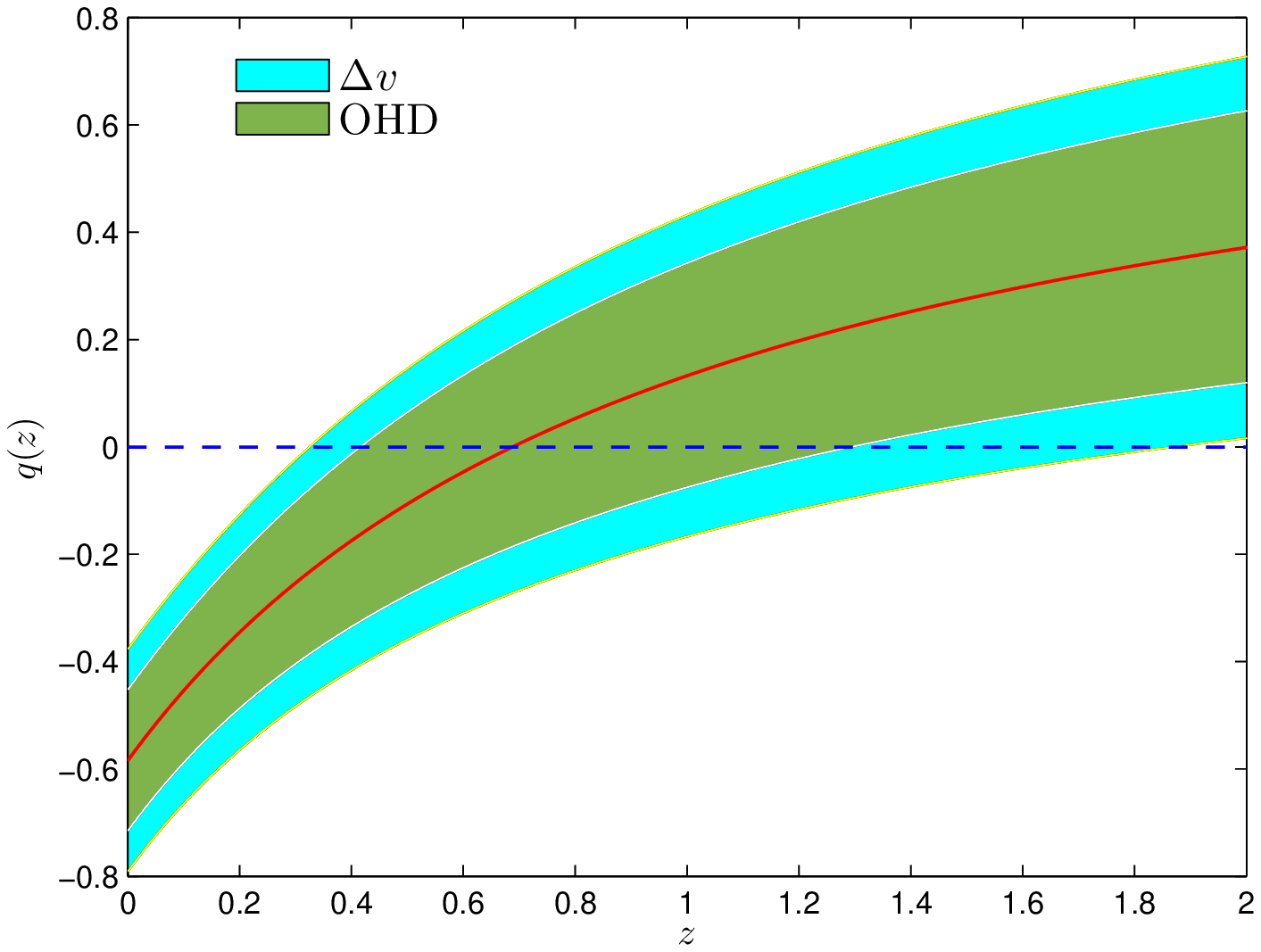}
\includegraphics[width=0.4\textwidth]{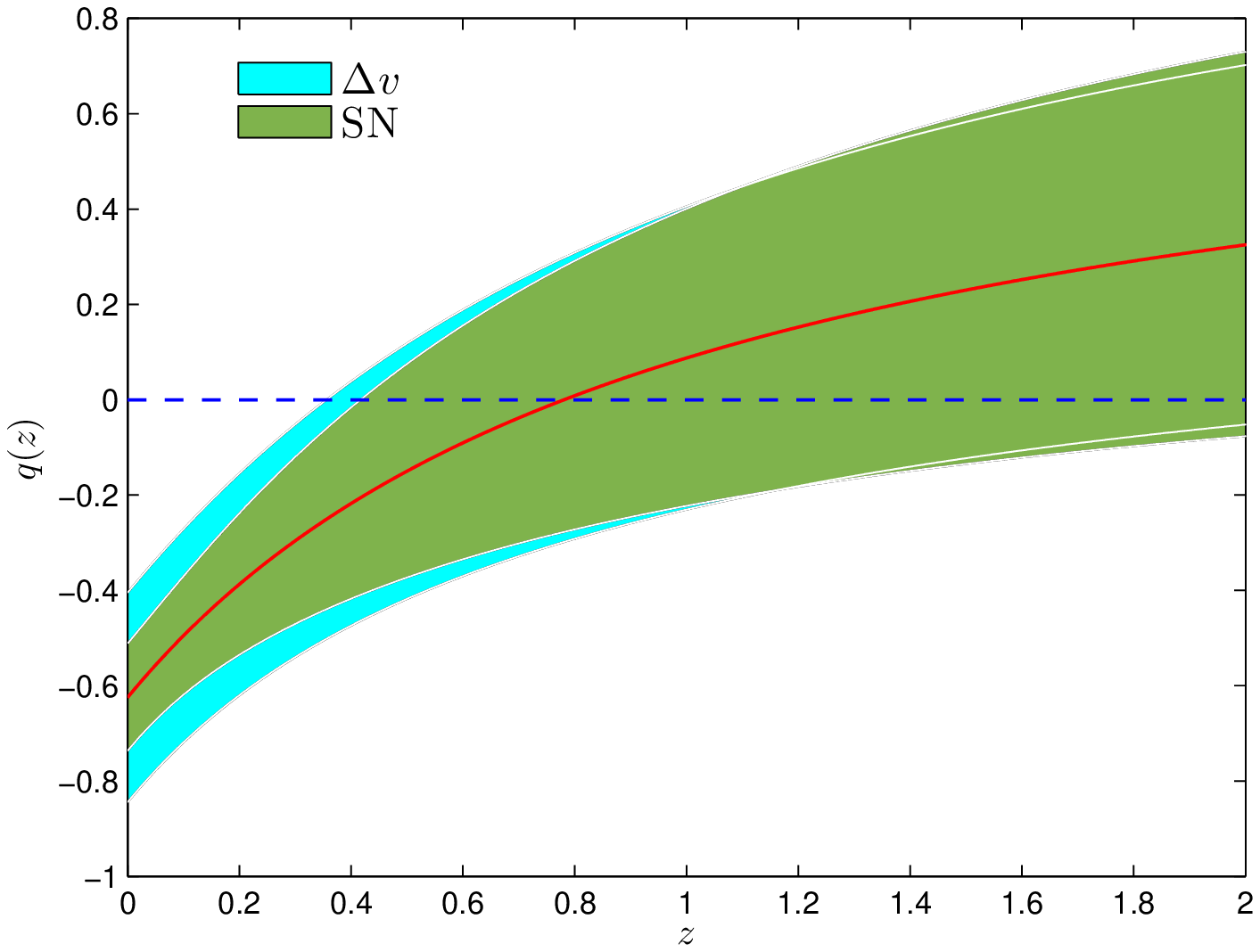}
    \caption{\label{q2reconstruction} Reconstruction of deceleration factor for model II:  $q(z)=q_0 + q_1 z/(1+z)$, using OHD, SNIa and $\Delta v$ with 10 years.}
\end{figure}

Another parametrization of considerable interest is
\citep{xu2007reconstruction}
    \be  \label{q model2}
    q(z)=q_0 + q_1 \frac{z}{1+z}.
    \ee
Difference from the above is that variation rate in this model is
not a constant but  $dq/dz= q_1/(1+z)^2$. Transition of this model
occurs at redshift $z_t=-q_0/(q_0+q_1)$. Physically, in the distant
past ($z \gg 0$), it leads to a constant $q(z)=q_0 + q_1$.

Performing operation introduced above, parameters fitted by OHD are
respectively $q_0=-0.5850^{+0.1308}_{-0.1318}$,
$q_1=1.4350^{+0.3266}_{-0.3230}$. A transition is therefore
estimated at $z_t=0.68^{+0.6054}_{-0.2718}$, later than that of
model I. It is evident that the dynamical model postpone the
accelerating expansion of universe. From Figure \ref{q2comparison},
pin-like constraint shape from $\Delta v$ leads to bigger
uncertainties of $q_0$ and $q_1$. Naturally, reconstructed $q(z)$
from $\Delta v$ is worse than that of OHD. Evidence for this can be
seen in top panel of Figure \ref{q2reconstruction}. Although
reconstruction from $\Delta v$ with 10 years is worse, we have
witnessed that its ability improves very fast with the increase of
observational time. The uncertainties of transition redshift for 10
years, 20 years and 30 years are estimated to be (+1.1822, -0.3629),
(+0.3693, -0.2165), (+0.2208, -0.1552), respectively. From 10 to 30
years, it reduces the uncertainties by 3--5 times. And 20 years are
enough to overwhelm the power of SNIa.

Cosmological fit by SNIa  provides
$q_0=-0.6250^{+0.1120}_{-0.1129}$, and
$q_1=1.4250^{+0.5836}_{-0.5785}$. Comparing with OHD, we note that
they imply a similar constraint on deceleration factor today, i.e.,
$-0.9 \lesssim q_0 \lesssim -0.3$. However, constraint on $q_1$ from
OHD is much narrower than that of SNIa. This result is also
demonstrated for model I. Reconstructed $q(z)$ in Figure
\ref{q2reconstruction} indicates transition from SNIa at
$z_t=0.78^{+2.2187}_{-0.3641}$, an earlier one than the constant
$dq/dz$ model, but a much looser upper uncertainty. Returning to our
focus $\Delta v$, we find that constraint on $q_0$ is terrible, but
a little better estimation on $q_1$ than that of SNIa. Finally,
$\Delta v$ gives $z_t=0.78^{+1.7461}_{-0.4250}$. If we extend the
observational time to 30 years, it can be reduced to
$z_t=0.78^{+0.2802}_{-0.1867}$. This estimation almost narrow by 8
times than that of SNIa.

\section{Conclusion and discussion}
\label{Conclusions}

We present some investigation on velocity drift $\Delta v$ based on
capability of the extremely stable and ultra high precision
spectrograph CODEX. Our survey is designed to untie the two
questions: How many future $\Delta v$ observational data could match
the constraint power of current OHD and SNIa? The second is, how
well future $\Delta v$ can provide information about the expansion
history? In which, constraints of OHD and SNIa are obtained using
the $\chi^2$ statistics, and constraints of $\Delta v$ are
implemented by the Fisher information matrix, according to the
modeled accuracy of spectroscopic $\Delta v$ in Equation
(\ref{velocity error}) by
\citet{pasquini2005codex,whitelock2006scientific}.

For the observational constraints in $\Lambda$CDM model, we obtain
that $\Delta v$ with 20 years could constrain the matter density
$\Omega_m$ to very high significance. The uncertainty $\Delta
\Omega_m$ could be determined at 0.0387, which reduces the
uncertainty by more than 42\% than available OHD and SNIa. Recently,
\citet{collaboration2013planck} report a high value of the matter
density parameter, $\Omega_m=0.315 \pm 0.017$. We also check that it
takes about 50 years to cover this power. Following the measurement
of the statistical FoM outlined in \citet{mortonson2010figures}, we
show that 21 future $\Delta v$ observations could be required to
satisfy the parameter constraint power of OHD, and 26 $\Delta v$ for
SNIa.

For the XCDM model, $\Delta v$ with 10 years is not enough to
precisely determine parameters. Purely statistic FoM indicate that
at least 12 years are required to match the power of OHD and SNIa on
$w$. Further investigation tells us that $\Delta v$ within 30 years
determine $\Omega_m$ almost with no degeneracy. Importantly, $\Delta
v$ with secular monitor could reduce the uncertainty of $w$ to very
high significance. So far, the most accurate observation on $w$
should attribute to the WMAP. In the report of WMAP9
\citep{hinshaw2012nine}, it issues $w=-1.122^{+0.068}_{-0.067}$ by
the combined WMAP+eCMB+BAO+$H_0$+SNe for non-flat XCDM model. For
the \citet{collaboration2013planck}, joint constraint from BAO and
CMB present $w=-1.13^{+0.13}_{-0.10}$. We check that individual
$\Delta v$ with 50 years provides an uncertainty $\Delta w=0.1098$,
which could be comparable with the Planck. In fact, the CMB-only
does not strongly constrain $w$. For example, result from WMAP9 is
$w>-2.1$ (95\%C.L.). We believe that $\Delta v$ for 50 years
combines with the first strong group must be far beyond current
results.

We also investigate two models of deceleration factor $q(z)$. For
the first model with constant $dq/dz$, OHD and SNIa give same
uncertainty of $q_0$. Estimations of transition redshift agree with
previous work, respectively. Results show that $\Delta v$ with only
10 years could provide much better constraint on them, especially
compared with SNIa. Moreover, constraint of $\Delta v$ on $q_0 -
q_1$ plane is almost orthogonal to that of OHD. Therefore, joint
constraints help us more to break the degeneracy. However, more
years for $\Delta v$ are needed to determine the variable $dq/dz$.
We check that $\Delta v$ for 30 years could reduce the transition
redshift to $z_t=0.68^{+0.2208}_{-0.1552}$ (basing on OHD) or
$z_t=0.78^{+0.2802}_{-0.1867}$ (basing on SNIa), which are more than
three times better than current estimations from OHD and SNIa.

Using the fashionable Fisher information matrix, we analysis the
power of $\Delta v$ with available OHD and SNIa. Fisher matrix could
provide relatively stable estimation on the uncertainty of
parameters. Our results quantitatively forecast that velocity drift
plays an important role on the cosmological research. From the
first-year maps of COBE in 1992 \citep{smoot1992structure} to the
up-to-date Planck, more than twenty years were sold. In the sense of
observational cost compared with current observations, $\Delta v$ is
a feasible measure with much better precision. On the other hand, it
could extend our knowledge of cosmic expansion into the deeper
redshift desert, where other probes are inaccessible. Moreover, it
is dynamical without any complicated calibration like SNIa.

Our results manifest the sensitivity of matter density $\Omega_m$ to
the $\Delta v$, which is in agreement with previous investigation
\citep{zhang2007exploring,jain2010constraints}. However, our work
reveals that how well $\Delta v$ numerically determines it for the
first time. Reconstructing deceleration factor indicates that
difference between these observations is constraint on variation
rate $q_1$. That is, constraint on variation rate of $q(z)$ from OHD
is much tighter than that of SNIa, no matter which kind of models.
While $\Delta v$ can exactly improve the determination of it. We
also note that OHD and SNIa give a similar estimation of
deceleration factor today, $-0.9 \lesssim q_0 \lesssim -0.3$.

Admittedly, there exists some deficiencies to advance. For instance,
our approaches are tentative and model-dependent. They are under the
assumption of $\Lambda$CDM model and XCDM model. For further
understanding, investigation for other models and observations are
necessary. For the important deceleration factor, it focus on the
second derivative of scale factor $a(t)$ with respect to cosmic
time, while redshift drift appearing in Equation (\ref{change
Hubble}) stands a variation of first derivative of $a(t)$. The
relation between them may also be investigated in our future work.
On the other hand, the FoM is statistical only and may not work well
for some situations. As explained by \citet{sendra2012supernova},
the FoM favors low correlation but behaves poorly when high
correlation is presented in the dark energy parameterization
considered. Furthermore, our analysis is based on the model by
\citet{pasquini2005codex,whitelock2006scientific} with 30 QSOs. We
anticipate more accurate conclusion could be done with the growth of
QSO's number.

\acknowledgments M.J.Zhang would like to thank Zhong-Xu Zhai and
Ying-Lin Wang for their valuable discussions. This work was
supported by the National Natural Science Foundation of China (Grant
Nos.11235003, 11175019, 11178007).

\bibliographystyle{hapj}

\bibliography{construction}

\begin{thebibliography}{67}
\expandafter\ifx\csname natexlab\endcsname\relax\def\natexlab#1{#1}\fi

\bibitem[{Acquaviva \& Gawiser(2010)}]{acquaviva2010falsify}
Acquaviva, V., \& Gawiser, E. 2010, Physical Review D, 82, 082001

\bibitem[{Albrecht {et~al.}(2009)Albrecht, Amendola, Bernstein, Clowe,
  Eisenstein, Guzzo, Hirata, Huterer, Kirshner, Kolb,
  {et~al.}}]{albrecht2009findings}
Albrecht, A. {et~al.} 2009, arXiv preprint arXiv:0901.0721

\bibitem[{Albrecht {et~al.}(2006)Albrecht, Bernstein, Cahn, Freedman, Hewitt,
  Hu, Huth, Kamionkowski, Kolb, Knox, {et~al.}}]{albrecht2006report}
------. 2006, arXiv preprint astro-ph/0609591

\bibitem[{Aldering {et~al.}(2002)Aldering, Akerlof, Amanullah, Astier,
  Barrelet, Bebek, Bergstrom, Bercovitz, Bernstein, Bester,
  {et~al.}}]{aldering2002overview}
Aldering, G. {et~al.} 2002, in SPIE Proceedings Series, Vol. 4835, 146--157

\bibitem[{Balbi \& Quercellini(2007)}]{balbi2007time}
Balbi, A., \& Quercellini, C. 2007, Monthly Notices of the Royal Astronomical
  Society, 382, 1623

\bibitem[{{Busca} {et~al.}(2013){Busca}, {Delubac}, {Rich},
  {et~al.}}]{2013A&A...552A..96B}
{Busca}, N.~G., {Delubac}, T., {Rich}, J., {et~al.} 2013, \aap, 552, A96,
  arXiv: 1211.2616

\bibitem[{Cunha \& Lima(2008)}]{cunha2008transition}
Cunha, J., \& Lima, J. 2008, Monthly Notices of the Royal Astronomical Society,
  390, 210

\bibitem[{Cunha(2009)}]{cunha2009kinematic}
Cunha, J.~V. 2009, Physical Review D, 79, 047301

\bibitem[{Darling(2012)}]{darling2012toward}
Darling, J. 2012, The Astrophysical Journal Letters, 761, L26

\bibitem[{Fairbairn \& Goobar(2006)}]{fairbairn2006supernova}
Fairbairn, M., \& Goobar, A. 2006, Physics Letters B, 642, 432

\bibitem[{Falk \& Schramm(1978)}]{falk1978limits}
Falk, S.~W., \& Schramm, D.~N. 1978, Physics Letters B, 79, 511

\bibitem[{Farooq \& Ratra(2013)}]{farooq2013hubble}
Farooq, O., \& Ratra, B. 2013, The Astrophysical Journal Letters, 766, L7

\bibitem[{Garnavich {et~al.}(1998)Garnavich, Jha, Challis, Clocchiatti,
  Diercks, Filippenko, Gilliland, Hogan, Kirshner, Leibundgut,
  {et~al.}}]{garnavich1998supernova}
Garnavich, P.~M. {et~al.} 1998, The Astrophysical Journal, 509, 74

\bibitem[{Gaztanaga {et~al.}(2009)Gaztanaga, Cabre, \&
  Hui}]{gaztanaga2009clustering}
Gaztanaga, E., Cabre, A., \& Hui, L. 2009, Monthly Notices of the Royal
  Astronomical Society, 399, 1663

\bibitem[{Hinshaw {et~al.}(2012)Hinshaw, Larson, Komatsu, Spergel, Bennett,
  Dunkley, Nolta, Halpern, Hill, Odegard, {et~al.}}]{hinshaw2012nine}
Hinshaw, G. {et~al.} 2012, arXiv:1212.5226

\bibitem[{Jain \& Jhingan(2010)}]{jain2010constraints}
Jain, D., \& Jhingan, S. 2010, Physics Letters B, 692, 219

\bibitem[{Jimenez \& Loeb(2008)}]{jimenez2008constraining}
Jimenez, R., \& Loeb, A. 2008, The Astrophysical Journal, 573, 37

\bibitem[{Jungman {et~al.}(1996)Jungman, Kamionkowski, Kosowsky, \&
  Spergel}]{jungman1996cosmological}
Jungman, G., Kamionkowski, M., Kosowsky, A., \& Spergel, D.~N. 1996, Physical
  Review D, 54, 1332

\bibitem[{Lin {et~al.}(2009)Lin, Cheng, Wang, Qiang, Ze-Long, Zhang, \&
  Wang}]{lin2009observational}
Lin, H., Cheng, H., Wang, X., Qiang, Y., Ze-Long, Y., Zhang, T.-J., \& Wang,
  B.-Q. 2009, Modern Physics Letters A, 24, 1699

\bibitem[{Linder(2006)}]{linder2006biased}
Linder, E.~V. 2006, Astroparticle Physics, 26, 102

\bibitem[{Liske {et~al.}(2008{\natexlab{a}})Liske, Grazian, Vanzella,
  Dessauges, Viel, Pasquini, Haehnelt, Cristiani, Pepe, Avila,
  {et~al.}}]{liske2008cosmic}
Liske, J. {et~al.} 2008{\natexlab{a}}, Monthly Notices of the Royal
  Astronomical Society, 386, 1192

\bibitem[{Liske {et~al.}(2008{\natexlab{b}})Liske, Grazian, Vanzella,
  Dessauges, Viel, Pasquini, Haehnelt, Cristiani, Pepe, Bonifacio,
  {et~al.}}]{liske2008elt}
------. 2008{\natexlab{b}}, The Messenger, 133, 10

\bibitem[{Liske {et~al.}(2009)Liske, Pasquini, Bonifacio, Bouchy, Carswell,
  Cristiani, Dessauges, D¡¯Odorico, D¡¯Odorico, Grazian,
  {et~al.}}]{liske2009espresso}
------. 2009, Science with the VLT in the ELT Era, 243

\bibitem[{Loeb(1998)}]{loeb1998direct}
Loeb, A. 1998, The Astrophysical Journal Letters, 499, L111

\bibitem[{Ma \& Zhang(2011)}]{ma2011power}
Ma, C., \& Zhang, T.-J. 2011, The Astrophysical Journal, 730, 74

\bibitem[{Maor {et~al.}(2001)Maor, Brustein, \&
  Steinhardt}]{maor2001limitations}
Maor, I., Brustein, R., \& Steinhardt, P.~J. 2001, Physical Review Letters, 86,
  6

\bibitem[{McVittie(1962)}]{mcvittie1962appendix}
McVittie, G. 1962, The Astrophysical Journal, 136, 334

\bibitem[{Mignone \& Bartelmann(2008)}]{mignone2008model}
Mignone, C., \& Bartelmann, M. 2008, Astronomy and Astrophysics, 481, 295

\bibitem[{Moresco {et~al.}(2012)Moresco, Cimatti, Jimenez, Pozzetti, Zamorani,
  Bolzonella, Dunlop, Lamareille, Mignoli, Pearce,
  {et~al.}}]{moresco2012improved}
Moresco, M. {et~al.} 2012, Journal of Cosmology and Astroparticle Physics,
  2012, 006

\bibitem[{Mortonson {et~al.}(2010)Mortonson, Huterer, \&
  Hu}]{mortonson2010figures}
Mortonson, M.~J., Huterer, D., \& Hu, W. 2010, Physical Review D, 82, 063004

\bibitem[{Nesseris \& Perivolaropoulos(2005)}]{nesseris2005comparison}
Nesseris, S., \& Perivolaropoulos, L. 2005, Physical Review D, 72, 123519

\bibitem[{Pasquini {et~al.}(2006{\natexlab{a}})}]{whitelock2006scientific}
Pasquini, {et~al.} 2006{\natexlab{a}}, in Whitelock, Patricia A and Dennefeld,
  Michel and Leibundgut, Bruno, Proc. IAU Symp. 232, The Scientific
  Requirements for ELT, 193--197

\bibitem[{Pasquini {et~al.}(2005)Pasquini, Cristiani, Dekker, Haehnelt, Molaro,
  Pepe, Avila, Delabre, D¡¯Odorico, Liske, {et~al.}}]{pasquini2005codex}
Pasquini, L. {et~al.} 2005, The Messenger, 122, 10

\bibitem[{Pasquini {et~al.}(2006{\natexlab{b}})Pasquini, Cristiani, Dekker,
  Haehnelt, Molaro, Shaver, Bonifacio, Borgani, D'Odorico, Vanzella,
  {et~al.}}]{pasquini2006codex}
------. 2006{\natexlab{b}}, Scientific Requirements for Extremely Large
  Telescopes, 232, 193

\bibitem[{Perivolaropoulos(2005)}]{perivolaropoulos2005constraints}
Perivolaropoulos, L. 2005, Physical Review D, 71, 063503

\bibitem[{Phillipps(1983)}]{phillipps1983limits}
Phillipps, S. 1983, Astrophysical Letters, 23, 145

\bibitem[{Pietro \& Claeskens(2003)}]{pietro2003future}
Pietro, E.~D., \& Claeskens, J.-F. 2003, Monthly Notices of the Royal
  Astronomical Society, 341, 1299

\bibitem[{Planck~Collaboration {et~al.}(2013)Planck~Collaboration, Aghanim,
  Armitage-Caplan, Arnaud, Ashdown, Atrio-Barandela, Aumont, Baccigalupi,
  Banday, {et~al.}}]{collaboration2013planck}
Planck~Collaboration, Ade, P. {et~al.} 2013, arXiv preprint arXiv:1303.5076

\bibitem[{Riess {et~al.}(2007)Riess, Filippenko, Challis, Clocchiatti, Diercks,
  Garnavich, Gilliland, Hogan, Jha, Kirshner, {et~al.}}]{riess1998supernova}
Riess, A.~G. {et~al.} 2007, The Astronomical Journal, 116, 1009

\bibitem[{Riess {et~al.}(2004)Riess, Strolger, Tonry, Casertano, Ferguson,
  Mobasher, Challis, Filippenko, Jha, Li, {et~al.}}]{riess2004type}
------. 2004, The Astrophysical Journal, 607, 665

\bibitem[{Sahni \& Starobinsky(2000)}]{sahni2000case}
Sahni, V., \& Starobinsky, A. 2000, International Journal of Modern Physics D,
  9, 373

\bibitem[{Samushia \& Ratra(2008)}]{samushia2008cosmological}
Samushia, L., \& Ratra, B. 2008, The Astrophysical Journal Letters, 650, L5

\bibitem[{Sandage(1962)}]{sandage1962change}
Sandage, A. 1962, The Astrophysical Journal, 136, 319

\bibitem[{Sendra \& Lazkoz(2012)}]{sendra2012supernova}
Sendra, I., \& Lazkoz, R. 2012, Monthly Notices of the Royal Astronomical
  Society, 422, 776

\bibitem[{Shafieloo {et~al.}(2006)Shafieloo, Alam, Sahni, \&
  Starobinsky}]{shafieloo2006smoothing}
Shafieloo, A., Alam, U., Sahni, V., \& Starobinsky, A.~A. 2006, Monthly Notices
  of the Royal Astronomical Society, 366, 1081

\bibitem[{Simon {et~al.}(2005)Simon, Verde, \& Jimenez}]{simon2005constraints}
Simon, J., Verde, L., \& Jimenez, R. 2005, Physical Review D, 71, 123001

\bibitem[{Smoot {et~al.}(1992)Smoot, Bennett, Kogut, Wright, Aymon, Boggess,
  Cheng, De~Amici, Gulkis, Hauser, {et~al.}}]{smoot1992structure}
Smoot, G.~F. {et~al.} 1992, The Astrophysical Journal, 396, L1

\bibitem[{Spergel {et~al.}(2003)Spergel, Verde, Peiris, Komatsu, Nolta,
  Bennett, Halpern, Hinshaw, Jarosik, Kogut, {et~al.}}]{spergel2003first}
Spergel, D.~N. {et~al.} 2003, The Astrophysical Journal Supplement Series, 148,
  175

\bibitem[{Starobinsky(1998)}]{starobinsky1998determine}
Starobinsky, A.~A. 1998, Journal of Experimental and Theoretical Physics
  Letters, 68, 757

\bibitem[{Stern {et~al.}(2010)Stern, Jimenez, Verde, Kamionkowski, \&
  Stanford}]{stern2010cosmic}
Stern, D., Jimenez, R., Verde, L., Kamionkowski, M., \& Stanford, S.~A. 2010,
  Journal of Cosmology and Astroparticle Physics, 2010, 008

\bibitem[{Suzuki {et~al.}(2012)Suzuki, Rubin, Lidman, Aldering, Amanullah,
  Barbary, Barrientos, Botyanszki, Brodwin, Connolly,
  {et~al.}}]{suzuki2012hubble}
Suzuki, N. {et~al.} 2012, The Astrophysical Journal, 746, 85

\bibitem[{Tegmark(1997)}]{tegmark1997measuring}
Tegmark, M. 1997, Physical Review Letters, 79, 3806

\bibitem[{Tegmark {et~al.}(2004)Tegmark, Strauss, Blanton, Abazajian, Dodelson,
  Sandvik, Wang, Weinberg, Zehavi, Bahcall, {et~al.}}]{tegmark2004cosmological}
Tegmark, M. {et~al.} 2004, Phys. Rev. D, 69, 103501

\bibitem[{Tegmark {et~al.}(1997)Tegmark, Taylor, \&
  Heavens}]{tegmark1997karhunen}
Tegmark, M., Taylor, A.~N., \& Heavens, A.~F. 1997, The Astrophysical Journal,
  480, 22

\bibitem[{Vogeley \& Szalay(1996)}]{vogeley1996eigenmode}
Vogeley, M.~S., \& Szalay, A.~S. 1996, The Astrophysical Journal, 465, 34

\bibitem[{Wang(2008)}]{wang2008figure}
Wang, Y. 2008, Physical Review D, 77, 123525

\bibitem[{Wang {et~al.}(2010)Wang, Percival, Cimatti, Mukherjee, Guzzo, Baugh,
  Carbone, Franzetti, Garilli, Geach, {et~al.}}]{wang2010designing}
Wang, Y. {et~al.} 2010, Monthly Notices of the Royal Astronomical Society, 409,
  737

\bibitem[{Wang \& Tegmark(2005)}]{wang2005uncorrelated}
Wang, Y., \& Tegmark, M. 2005, Physical Review D, 71, 103513

\bibitem[{Wei(2010)}]{wei2010observational}
Wei, H. 2010, Journal of Cosmology and Astroparticle Physics, 2010, 020

\bibitem[{Wei {et~al.}(2007)Wei, Tang, \& Zhang}]{wei2007reconstruction}
Wei, H., Tang, N., \& Zhang, S.~N. 2007, Physical Review D, 75, 043009

\bibitem[{Wu {et~al.}(2012)Wu, Ma, \& Zhang}]{wu2012reconstructing}
Wu, C.-J., Ma, C., \& Zhang, T.-J. 2012, The Astrophysical Journal, 753, 97

\bibitem[{Xu {et~al.}(2007)Xu, Zhang, Chang, \& Liu}]{xu2007reconstruction}
Xu, L., Zhang, C., Chang, B., \& Liu, H. 2007, arXiv preprint astro-ph/0701519

\bibitem[{Yoo {et~al.}(2011)Yoo, Kai, \& Nakao}]{yoo2011redshift}
Yoo, C.-M., Kai, T., \& Nakao, K.-i. 2011, Physical Review D, 83, 043527

\bibitem[{Zhai {et~al.}(2011)Zhai, Zhang, \& Liu}]{zhai2011constraints}
Zhai, Z.-X., Zhang, T.-J., \& Liu, W.-B. 2011, Journal of Cosmology and
  Astroparticle Physics, 2011, 019

\bibitem[{Zhang {et~al.}(2007)Zhang, Zhong, Zhu, \& He}]{zhang2007exploring}
Zhang, H., Zhong, W., Zhu, Z.-H., \& He, S. 2007, Physical Review D, 76, 123508

\bibitem[{Zhang {et~al.}(2010{\natexlab{a}})Zhang, Zhang, \&
  Zhang}]{zhang2010sandage}
Zhang, J., Zhang, L., \& Zhang, X. 2010{\natexlab{a}}, Physics Letters B, 691,
  11

\bibitem[{Zhang {et~al.}(2010{\natexlab{b}})Zhang, Ma, \&
  Lan}]{zhang2010constraints}
Zhang, T.-J., Ma, C., \& Lan, T. 2010{\natexlab{b}}, Advances in Astronomy,
  2010, 81

\end{thebibliography}

\end{document}